\documentclass[12pt]{iopart}

\usepackage{txfonts}
\usepackage{subcaption}
\usepackage{graphicx}
\usepackage{xcolor}
\usepackage{dcolumn}
\usepackage{hyperref}
\usepackage{longtable}

\usepackage{amssymb}
\usepackage{xspace}
\usepackage{soul}
\usepackage[mathlines]{lineno}
\usepackage{aas_macros}
\usepackage{orcid}
\usepackage[normalem]{ulem}

\newcommand{\chicrit}{\ensuremath{\chi_{\rm crit}}\xspace}
\newcommand{\refl}{\ensuremath{\mathcal{R}}\xspace}
\newcommand{\mos}{\ensuremath{m_{\rm 1,s}}\xspace}
\newcommand{\mts}{\ensuremath{m_{\rm 2,s}}\xspace}
\newcommand{\com}{\ensuremath{\chi_{\rm 1,M}}\xspace}
\newcommand{\ctm}{\ensuremath{\chi_{\rm 2,M}}\xspace}
\newcommand{\Ngw}{\ensuremath{N_{\rm GW}}\xspace}

\newcommand{\Ncbc}{\ensuremath{N_{\rm CBC}}\xspace}
\newcommand{\Nexp}{\ensuremath{N_{\rm exp}}\xspace}
\newcommand{\Pdet}{\ensuremath{P_{\rm det}}\xspace}

\newcommand{\feco}{\ensuremath{f_{\rm ECO}}\xspace}
\newcommand{\leps}{\ensuremath{\log_{10}(\epsilon)}\xspace}
\newcommand{\bbh}{\texttt{BBH}\xspace}
\newcommand{\eco}{\texttt{ECO}\xspace}
\newcommand{\mixture}{\texttt{Mixture}\xspace}

\newcommand{\Tobs}{\ensuremath{T_{\rm obs}}\xspace}
\newcommand{\xdata}{\ensuremath{\{x\}}\xspace}
\newcommand{\nn}{\nonumber}

\newcommand{\de}{{\rm d}}

\newcommand{\icarogw}{\textsc{icarogw}\xspace} 

\usepackage{iopams}  
\begin{document}

\title[Limits on the existence of totally reflective exotic compact objects]{Limits on the existence of totally reflective exotic compact objects with current and future gravitational-wave detectors}

\author{S. Mastrogiovanni$^1$ \orcidlink{0000-0003-1606-4183}, Elisa Maggio$^{2}$ \orcidlink{0000-0002-1960-8185}, Adriano Frattale Mascioli$^{1,3}$ \orcidlink{0000-0002-0155-3833}} 
\address{$^1$INFN, Sezione di Roma, I-00185 Roma, Italy
\\$^2$Max Planck Institute for Gravitational Physics (Albert Einstein Institute), Am M\"uhlenberg 1, Potsdam 14476, Germany
\\$^3$Dipartimento di Fisica, Università di Roma “Sapienza”, Piazzale A. Moro 5, I-00185, Roma, Italy}

\ead{mastrosi@roma1.infn.it}
\vspace{10pt}

\begin{abstract}
Exotic compact objects (ECOs) are a theorized class of compact objects that solve the paradoxes of black holes
by replacing the event horizon with a physical surface located at $r=r_+(1+\epsilon)$ from the would-be horizon at $r_+$. Spinning horizonless objects are prone to the ergoregion instability, which would prevent their existence if their spin is higher than a critical threshold. In this paper, we set upper limits on the existence of a population of merging ECOs from the spin distribution of the population of compact binary coalescences (CBCs) detected by the LIGO, Virgo and KAGRA collaboration. Using spin measurements from 104 compact objects, we find that if ECOs have $\epsilon \in [10^{-42}-10^{-3}]$ and their surface is totally reflective, the population of CBCs cannot be composed (at 90\% credible level) by more than 71\% (59\%) of ECOs for polar (axial) perturbations. If we restrict the ECOs to be ultracompact ($\epsilon<10^{-30}$), at 90\% credible level, ECOs cannot compose more than 28\% and 25\% of the CBC population for polar and axial perturbations. The constraints from current data are a factor of two more precise than the ones obtained from a non-detection of a stochastic GW background due to spin loss. We also study how next generation gravitational-wave detectors, such as the Einstein Telescope, can constrain the ECO population. We find that 1 day of data taking would be enough to constrain the fraction of ECOs to be lower than 20\% for $\epsilon \in [10^{-42}-10^{-3}]$.
\end{abstract}

%
%
%
\maketitle
%
%

\section{Introduction}

Black holes (BHs) are arguably the most compact and fascinating objects in the universe. Although much progress has been made in describing their phenomenology in the electromagnetic and gravitational sectors~\cite{GRAVITY:2020gka, EventHorizonTelescope:2019pgp, EventHorizonTelescope:2022xqj, LIGOScientific:2016lio, LIGOScientific:2019fpa, LIGOScientific:2020tif, LIGOScientific:2021sio}, there are still several theoretical and observational properties that are not well understood. From a theoretical point of view, BHs are endowed with an event horizon beyond which closed timelike curves can violate causality and are characterized by an information-loss paradox due to the emission of Hawking radiation~\cite{1974Natur.248...30H,Mathur_2009}. Furthermore, the BH interior hides a curvature singularity in which general relativity pathologically breaks down. From an observational point of view, the LIGO \cite{LIGOScientific:2014pky}, Virgo \cite{VIRGO:2014yos}, and KAGRA \cite{Aso:2013eba} (LVK) collaboration has observed gravitational waves (GWs) from compact binary coalescences (CBCs) whose masses are in expected mass gap between BHs and neutron stars (NSs)~\cite{LIGOScientific:2018mvr,LIGOScientific:2020ibl,LIGOScientific:2021usb,KAGRA:2021vkt,KAGRA:2021duu}. In particular, the lighter compact object of GW190814 \cite{2020ApJ...896L..44A} has been found in the expected mass gap between NSs and BHs and
the primary mass of GW190521 lies within the gap produced by pair-instability supernova processes \cite{LIGOScientific:2020iuh,2020ApJ...900L..13A}.

Exotic Compact Objects (ECOs) are theorized possible solutions to the information-loss paradox and the singularity problem \cite{2016JCAP...10..001G}. 
ECO models are constructed to preserve the phenomenology observed at the exterior of BHs, while changing the physics at the horizon scale and the internal structure. For this reason, ECOs are typically referred to as ``BH mimickers'' \cite{Lemos:2008cv, Cardoso:2007az} and try to solve the BH paradoxes by replacing the event horizon with a surface with reflectivity \refl at the radius $r_0=r_{\rm H}(1+\epsilon)$ outside the would-be event horizon $r_{\rm H}$. Depending on the value of $\epsilon$ (which is related to the compactness of the object) and \refl, ECOs are classified into different families; see \cite{Cardoso:2019rvt, 2021hgwa.bookE..29M} for recent reviews.
ECO models introduce a new type of phenomenology that can be tested with astrophysical observations, for example, for the tidal heating~\cite{Datta:2019epe,Maggio:2021uge,Mukherjee:2022wws} and tidal deformability~\cite{Cardoso:2017cfl,Chakraborty:2023zed} in the inspiral, the modified quasinormal mode spectrum in the ringdown, and the emission of GW echoes~\cite{Cardoso:2016rao,Cardoso:2016oxy,Maggio:2020jml}. 

In this paper, we focus on how the ergoregion instability in spinning horizonless compact objects~\cite{Friedman:1978ygc} can leave a fingerprint in the population distribution of CBCs.
The former is an instability that develops in any spacetime with an ergoregion but without a horizon: since negative-energy states can exist inside the ergoregion, it is energetically favorable to cascade towards more negative states leading to a runaway instability~\cite{Penrose:1969pc,Brito:2015oca}.
The instability is triggered when the spin of the horizonless object is beyond a critical value of the spin, which is determined by the ECO's compactness and reflectivity \cite{2017PhRvD..96j4047M,2019PhRvD..99f4007M}. 
The object would lose energy and angular momentum via GWs until the spin reaches the critical threshold for the  instability process.
The ergoregion instability was analyzed in models of uniform-density stars~\cite{1978RSPSA.364..211C,10.1093/mnras/282.2.580,Kokkotas:2002sf}, gravastars~\cite{Chirenti:2008pf}, boson stars~\cite{Cardoso:2007az}, and superspinars~\cite{Cardoso:2008kj,Pani:2010jz}.
The only way to prevent such an infinite cascade is by absorbing the negative-energy states efficiently with dissipation mechanisms within the object~\cite{2019PhRvD..99f4007M}.

If perfectly reflecting ECOs exist, we would expect to observe a population of compact objects with spins lower than the critical threshold predicted by an ECO model.  We note that the spin magnitude distribution of compact objects has already been used as a possible smoking gun for the presence of ultra-light boson clouds around spinning BHs \cite{Ng:2019jsx,Ng:2020ruv}. Ultra-light boson clouds around spinning BHs can trigger a superradiance instability, whose time scales and conditions depend on the BH and boson masses~\cite{Brito:2015oca,Brito:2017wnc}. While the ergoregion and superradiance instabilities share many common aspects, such as the correspondence between the
minimum absorption coeﬃcient and the maximum superradiant amplification~\cite{2017PhRvD..96j4047M,2019PhRvD..99f4007M}, they are two different problems in theoretical physics.

A population of ECOs can compose part of the CBCs we detect with GWs. 
In this paper, we study the spin distribution of the compact objects detected with GWs to probe the existence of a hidden population of ECOs. The paper is organized as follows. 
In Sec.~\ref{sec:2}, we provide an introduction to the ergoregion instability and its dynamical timescales. In Sec.~\ref{sec:3}, we construct a spin magnitude population model to describe a CBC population of ECOs and BHs. In Sec.~\ref{sec:4}, we discuss the constraints from 104 compact objects reported in the third Gravitational Wave Transient Catalog (GWTC-3) \cite{KAGRA:2021vkt} and, in Sec.~\ref{sec:5}, we report projections for future ground-based GW detectors. Finally, in Sec.~\ref{sec:6}, we draw our conclusions.

\section{Ergoregion instability for totally reflective exotic compact objects}
\label{sec:2}

ECO models are constructed by replacing the BH horizon located at $r=r_+$ with a perfectly reflecting surface at $r=r_+(1+\epsilon)$ where $\epsilon>0$ and $\epsilon \ll 1$. Scalar, vector (electromagnetic) and tensor (gravitational) perturbations can be trapped between the ECO's surface and the photon-sphere barrier if the object is sufficiently compact.
While trapped, the perturbations undergo a ``bounce-and-amplify'' process between the ECO surface and the photon sphere. As in a Fabry-Perot cavity, the frequency of the trapped modes is set by the cavity's width, which depends on the ECO's mass and spin, whereas the damping time of the modes depends on the surface's absorption.  Here we focus on perfectly reflecting ECOs for which $\left|\refl(\omega)\right|^2 = 1$. In the remainder of the paper, we drop the dependencies of the various quantities from the reflectivity in order to simplify the notation.

Depending on the type of perturbation, it is possible to calculate analytically the real and imaginary parts, $\omega_R$ and $\omega_I$, for trapped modes \footnote{With some caveats in the small-frequency regime for which we defer the reader to Ref.~\cite{2019PhRvD..99f4007M} for vector and tensor perturbations and to Ref.~\cite{2017PhRvD..96j4047M} for the scalar ones.}. If the imaginary part of the frequency of the trapped modes $\omega_I$ changes sign, the ergoregion instability is triggered after a time scale $\tau_{\rm inst}$. Then, the object starts to lose rotational energy in GWs eventually quenching the ergoregion instability~\cite{Barausse:2018vdb}. Another time scale relevant to this process is the spin-down timescale $\tau_{\rm sd}$ when the object is spanned down to the critical spin for which the ergoregion instability is quenched.

\begin{figure}[t]
    \centering
    \includegraphics{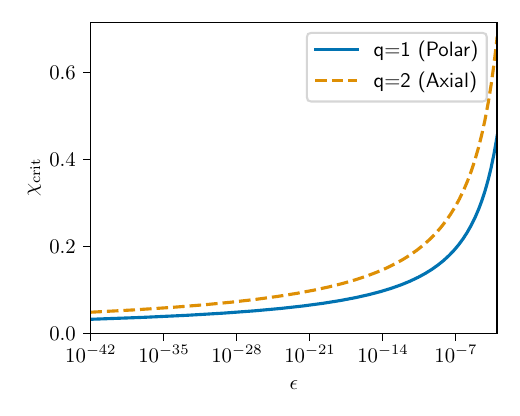}
    \caption{Critical spin for the ergoregion instability as a function of $\epsilon$ for a reflective ECO whose radius is located at $r_0 = r_+ (1+\epsilon)$ where $r_+$ is the BH horizon. The blue solid line represents \chicrit for polar perturbations, while the orange dashed line is for axial perturbations.}
    \label{fig:chicrit}
\end{figure}

As we expect ECOs to be formed by direct gravitational collapse or merger of two ECOs, in this paper we focus on the ergoregion instability triggered by gravitational perturbations. For gravitational perturbations and perfectly reflecting ECOs, the critical spin magnitude\footnote{We indicate with spin magnitude the mass-renormalized spin, $\chi \in [0,1]$.} at which the ergoregion instability is triggered, for $\epsilon \ll 1$, is given by \cite{2019PhRvD..99f4007M}
\begin{equation}
    \chicrit(\epsilon) = \frac{\pi (1+q)}{m|\log_{10}\epsilon|},
    \label{eq:chicrit}
\end{equation}
where $q=1,2$ for polar and axial perturbations and $m$ is the perturbation azimuthal number. In this paper, we focus on the dominant quadrupolar perturbation ($l=m=2$). Fig.~\ref{fig:chicrit} shows the critical spin as a function of $\epsilon$ for polar and axial perturbations. As inferred from the plot, the critical spin is typically $\chi_{\rm{crit}} \sim 0.05$ for ultracompact ECOs with $\epsilon \sim 10^{-40}$ and is approximately $\chi_{\rm{crit}} \sim 0.6$ for $\epsilon \sim 10^{-3}$, which is a reasonable upper limit for small deviations from the would-be horizon~\cite{Maggio:2020jml}. 

The instability time scale is defined as $\tau_{\rm{inst}}=1/|\omega_I|$. For a compact object with $\epsilon=10^{-10}$ and spin $\chi=0.7$, the instability time scale of the $l=m=2$ mode is~\cite{2021hgwa.bookE..29M}
\begin{equation}
    \tau_{\rm{inst}} \in (5,7) \left(\frac{M}{10 M_\odot}\right) \ \rm{s},
\end{equation}
where the lower (upper) bound is for polar (axial) perturbations. The ergoregion instability is expected to be triggered at times that are not cosmologically relevant. 
In fact, a $\tau_{\rm inst} \sim 1 \ {\rm Myr}$ would require $M\omega_I \sim 10^{-19}$, which is a value reached only when $|\chi-\chi_{\rm crit}| \ll 1$~\cite{2019PhRvD..99f4007M}. 
The ergoregion instability acts on a timescale which is short compared to the accretion timescale of astrophysical BHs, $\tau_{\rm{Salpeter}} \sim 4 \times 10^7 \ \rm{yr}$, and longer than the decay time of the BH ringdown, $\tau_{\rm{ringdown}} \sim 0.5 \ \rm{ms}$ for a $10 \ M_\odot$ object.
As such, we can safely assume that regardless of $\epsilon$, initial mass and spin of the ECO, the ergoregion instability will be triggered if $\chi>\chi_{\rm crit}$ on a time that is not cosmologically relevant. 
The object would then lose angular momentum in GWs over a time scale $\tau_{\rm{sd}}$ until the stability condition, $\chi=\chi_{\rm crit}$, is satisfied. A qualitative estimate of the spin-down time can be obtained using the quasi-adiabatic approximation \cite{Barausse:2018vdb} for times $t \gg \tau_{\rm inst}$. This condition is always satisfied for cosmological times and it allows us to describe the extraction of the orbital angular momentum, $J$, and the object's energy, $E$, as
\begin{eqnarray}
\dot{J}&=&-m \frac{\dot{E}_0}{\omega_R},\\
\dot{E}&=&\dot{E}_0 e^{\omega_I t},
\end{eqnarray}
where $\dot{E}_0$ encodes the initial perturbation of the unstable system. The equations above indicate that for times $t \gg \tau_{\rm inst}$, the extraction of the rotational energy will be exponential. After this phase, the extraction of the rotational energy will become inefficient and $\chi$ will approach $\chi_{\rm crit}$ asymptotically for $t \rightarrow \infty$. As a consequence, we can assume that in a timescale that is not cosmologically relevant, the ECO's spin will rapidly approach the value $\chi \sim \chi_{\rm crit}$, and then it will continue emitting a monochromatic (and weak) GW signal can be detected as a stochastic background \cite{Barausse:2018vdb}.

\section{Population and statical modelling}
\label{sec:3}

In the previous section, we have discussed that, for a given compactness, reflective ECOs display a critical spin above which the ergoregion instability is triggered. If an ECO is subjected to the ergoregion instability, it  spins down to a value of $\chi \sim \chicrit$ 
emitting a weak GW signal detectable as a stochastic background. Here, we do not focus on the stochastic background, but on the fact that ECOs in compact binaries are expected to have $\chi \leq \chicrit$.

\begin{figure}[t]
    \centering
    \includegraphics[scale=0.3]{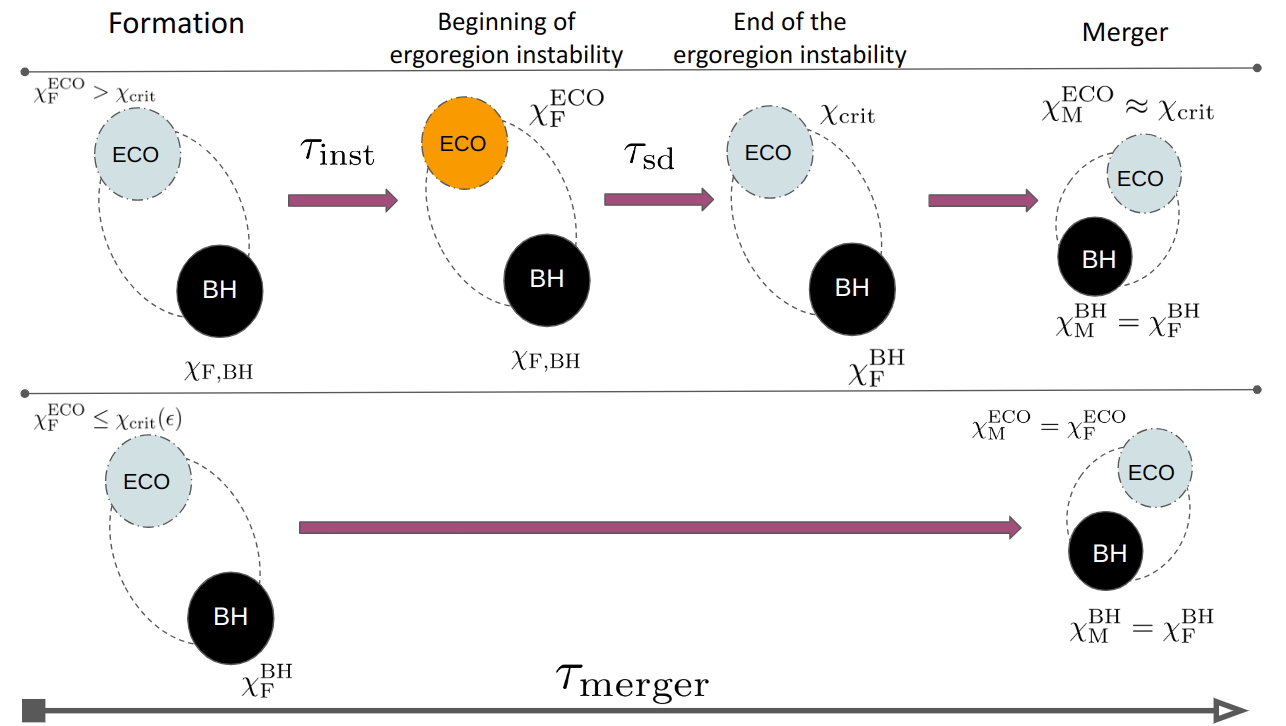}
    \caption{Conceptual model to construct the spin distribution at merger of CBCs composed by an ECO and a classical BH. The figure displays two possible cases. \textit{Top panel:} The binary is formed, the ECO is subjected to the ergoregion instability and modifies its spin before the merger. \textit{Bottom panel:} The binary is formed, the ergoregion instability is not triggered and the ECO's spin is conserved until the merger.}
    \label{fig:draw}
\end{figure}

If the population of the detected CBCs is partially composed of ECOs, we expect to observe non-trivial features in the spin distribution of the population. In Fig.~\ref{fig:draw}, we provide a conceptual representation of our population model. Let us consider a binary formed by an ECO and a classical BH.
There are two possible scenarios for the evolution of the binary. In the first case, represented in the bottom panel of Fig.~\ref{fig:draw}, the ECO at formation has a spin  $\chi^{\rm ECO}_{\rm F}\leq \chicrit$, and after a time $\tau_{\rm merger}$ it merges with the BH.
If the binary is isolated during the coalescence and non-precessing, we expect the spins at the merger to be $\chi^{\rm ECO}_{\rm M}=\chi^{\rm ECO}_{\rm F}$ and $\chi_{\rm M}^{\rm BH}=\chi_{\rm F}^{\rm BH}$. 
In the second case, represented in the top panel of Fig.~\ref{fig:draw}, the ECO at formation has a spin $\chi^{\rm ECO}_{\rm F} > \chicrit$. After a time $\tau_{\rm inst}$, the ergoregion instability is triggered and the ECO begins to spin down from $\chi^{\rm ECO}_{\rm F}$ to approximately $\chicrit$ in the spin-down time $\tau_{\rm sd}$. Close to the merger, the ECO has a spin $\chi^{\rm ECO}_{\rm M} \approx \chicrit$, while the classical BH has a spin $\chi^{\rm BH}_{\rm M}=\chi^{\rm BH}_{\rm F}$. 
If both the objects in the binary are ECOs, they will still follow the dynamics described in the panels of Fig.~\ref{fig:draw}. Whereas, if the two objects are BHs, they will have a spin distribution at merger equal to the spin distribution at formation.

\subsection{Spin magnitude distributions} \label{sec:pop_model}

We now formally construct the spin distribution at the merger of CBCs. When measuring the spin magnitudes of a CBC, we do not know a priori if the two objects are ECOs or BHs. As such, we need to marginilize upon their nature. The differential number of CBCs merging with spins $\chi_{\rm 1,M}$ and $\chi_{\rm 2,M}$, where the subscripts $1,2$ indicate the two objects in the binary,  is given by
\begin{eqnarray}
    \frac{\de N_{\rm CBC}}{\de \chi_{\rm 1,M} \ \de \chi_{\rm 2,M}}= \sum^{\rm BH,ECO}_{\rm A} \sum^{\rm BH,ECO}_{\rm B} \frac{\de N_{\rm CBC}}{\de \chi^{\rm A}_{\rm 1,M} \ \de \chi^{\rm B}_{\rm 2,M}}p(A)p(B) \,,
   \label{eq:ratespin}
\end{eqnarray}
where the indices $A,B$ indicate the nature of the objects and $p(A),p(B)$ are the probabilities that the objects are formed as BHs or ECOs. A simple parametrization for this probability is that ECOs compose a certain fraction $f_{\rm ECO}$ of the overall compact objects population, namely $p(\rm ECO)=f_{\rm ECO}$ and $p(\rm BH)=1-f_{\rm ECO}$.
The rate term in the right hand side of Eq.~(\ref{eq:ratespin}) can be written in terms of the spins at formation of the compact objects as
\begin{eqnarray}
    \frac{\de N_{\rm CBC}}{\de \chi^{\rm A}_{\rm 1,M} \ \de \chi^{\rm B}_{\rm 2,M}}= 
    \int \de \chi^{\rm A}_{\rm 1,F} \ \de \chi^{\rm B}_{\rm 2,F} \ \frac{\de N_{\rm CBC}}{\de \chi^{\rm A}_{\rm 1,F} \ \de \chi^{\rm B}_{\rm 2,F}} p\left(\chi^{\rm A}_{\rm 1,M}, \ \chi^{\rm B}_{\rm 2,M} \ | \ \chi^{\rm A}_{\rm 1,F}, \chi^{\rm B}_{\rm 2,F},A,B\right) \,,
    \label{eq:ratespin1}
\end{eqnarray}
where 
\begin{equation}
    \frac{\de N_{\rm CBC}}{\de \chi^{\rm A}_{\rm 1,F} \ \de \chi^{\rm B}_{\rm 2,F}}=N_{\rm CBC} \ \beta_{\rm F}\left(\chi^{\rm A}_{\rm 1,F}|\alpha,\beta\right) \ \beta_{\rm F}\left(\chi_{\rm 2,F}^{\rm B}|\alpha,\beta\right)
    \label{eq:form}
\end{equation}
is the CBC spin distribution at formation with \Ncbc the number of CBCs in the universe. 
We assume that both ECOs and classical BHs are formed with spins drawn from a common distribution that we describe as a $beta$ distribution governed by two population parameters $\alpha,\beta$. The motivation for assuming a common spin distribution for the formation of the two compact objects is that the astrophysical formation mechanisms and the angular momentum conservation laws leading to the formation of BHs and ECOs could be similar. 
The other relevant quantity in the right-hand side of Eq.~(\ref{eq:ratespin1}) is the probability that the two objects will merge with spins $\chi^{\rm A}_{\rm 1,M},\chi^{\rm B}_{\rm 1,M}$ while they were formed with  spins $\chi^{\rm A}_{\rm 1,F},\chi^{\rm B}_{\rm 1,F}$. According to our model in Fig.~\ref{fig:draw}, this probability only depends on whether the object is a BH or an ECO and therefore it can be factorized as 
\begin{equation}
    p\left(\chi^{\rm A}_{\rm 1,M}, \chi^{\rm B}_{\rm 2,M} \ | \ \chi^{\rm A}_{\rm 1,F}, \chi^{\rm B}_{\rm 2,F}, A, B\right)=p\left(\chi^{\rm A}_{\rm 1,M} \ | \ \chi^{\rm A}_{\rm 1,F},A\right) \ p\left(\chi^{\rm B}_{\rm 2,M} \ | \ \chi^{\rm B}_{\rm 2,F},B\right).
    \label{eq:pppd}
\end{equation}
Since the formation spins of the two objects, in Eq.~(\ref{eq:form}), and their evolution to the merger spins, in Eq.~(\ref{eq:pppd}), are independent between the two bodies, we can rewrite Eq.~(\ref{eq:ratespin}) as
\begin{equation}
\frac{\de N_{\rm CBC}}{\de \chi^{\rm A}_{\rm 1,M} \ \de \chi^{\rm B}_{\rm 2,M}}= N_{\rm CBC} \ p_{\rm pop}\left(\chi^{\rm A}_{\rm 1,M} \ | \ \alpha,\beta,A\right) \ p_{\rm pop}\left(\chi^{\rm B}_{\rm 2,M} \ | \ \alpha,\beta,B\right) \,,
\label{eq:ciao}
\end{equation}
having defined 
\begin{eqnarray}
   p_{\rm pop}\left(\chi^{\rm A}_{\rm 1,M} \ | \ \alpha,\beta,A\right) &=& \int  \de \chi^{\rm A}_{\rm 1,F} \ p\left(\chi^{\rm A}_{\rm 1,M} \ | \ \chi^{\rm A}_{\rm 1,F},A\right) \ \beta_{\rm F}\left(\chi^{\rm A}_{\rm 1,F} \ | \ \alpha,\beta\right) \,,\label{eq:mergerspin1} \\ 
   p_{\rm pop}\left(\chi^{\rm B}_{\rm 2,M} \ | \ \alpha,\beta,B\right) &=& \int \de \chi^{\rm B}_{\rm 2,F} \ p\left(\chi^{\rm B}_{\rm 2,M} \ | \ \chi^{\rm B}_{\rm 2,F},B\right) \ \beta_{\rm F}\left(\chi^{\rm B}_{\rm 2,F} \ | \ \alpha,\beta\right)
    \label{eq:mergerspin}.
\end{eqnarray}
Finally, replacing Eq.~(\ref{eq:ciao}) into Eq.~(\ref{eq:ratespin}) and recalling that $p(\rm ECO)=\feco$ and $p(\rm BH)=1-\feco$, we can define an overall spin distribution for the two objects at merger,
\begin{eqnarray}
\frac{\de N_{\rm CBC}}{\de \chi_{\rm 1,M} \ \de \chi_{\rm 2,M}}&=& \Ncbc \ p_{\rm pop}\left(\chi_{\rm 1,M}|\alpha,\beta\right) \ p_{\rm pop}\left(\chi_{\rm 2,M}|\alpha,\beta\right)  \nn \\ &=& \nn \Ncbc \ \Big[(1-\feco) \ p_{\rm pop}\left(\chi^{\rm BH}_{\rm 1,M} | \alpha,\beta,{\rm BH}\right)+ \\ \nn &&\feco \ p_{\rm pop}\left(\chi^{\rm ECO}_{\rm 1,M} | \alpha,\beta,{\rm ECO}\right)\Big] \Big[(1-\feco) \ p_{\rm pop}\left(\chi^{\rm BH}_{\rm 2,M}| \alpha,\beta,{\rm BH}\right)+ \\ &&\feco \ p_{\rm pop}\left(\chi^{\rm ECO}_{\rm 2,M}| \alpha,\beta,{\rm ECO}\right)\Big] \,.
\label{eq:ffff}
\end{eqnarray}

The only two missing terms to explicitly calculate the spin distribution of the CBC population are the analytical form of Eqs.~(\ref{eq:mergerspin1})-(\ref{eq:mergerspin}). Below, we will compute it only for the primary component as for the secondary is equivalent. 
If the object is a classical BH, the spin at the merger is assumed to be equal to the spin at the formation and 
\begin{equation}
    p\left(\chi^{\rm BH}_{\rm 1,M} \ | \ \chi^{\rm BH}_{\rm 1,F},{\rm BH}\right) = \delta\left(\chi^{\rm BH}_{\rm 1,M}-\chi^{\rm BH}_{\rm 1,F}\right),
\end{equation}
where $\delta$ is a Dirac's delta distribution. It follows that the spin distribution at merger for BHs is equal to the spin distribution at formation, namely
\begin{equation}
    p_{\rm pop}\left(\chi^{\rm BH}_{\rm 1,M} \ | \ \alpha,\beta,{\rm BH}\right) = \beta_{\rm F}\left(\chi^{\rm BH}_{\rm 1,M} \ | \ \alpha,\beta\right).
    \label{eq:spinfinal1}
\end{equation}
If the object is an ECO, there are two possible scenarios as depicted in Fig.~\ref{fig:draw}. 
If the ECO is formed with $\chi^{\rm ECO}_{\rm 1, F} \leq \chicrit(\epsilon)$, the spin at the merger is equal to the spin at formation. While if the ECO is formed with $\chi^{\rm ECO}_{\rm 1, F} > \chicrit(\epsilon)$, it undergoes a spin-down process  that rapidly reduces the spin until it reaches approximately \chicrit. As we expect the spin to asymptotically approach \chicrit, we assume an agnostic gaussian distribution for the final value of the ECO's spin.
For the ECO, the probability of having a merger spin given a formation spin is
\begin{eqnarray}
p\left(\chi^{\rm ECO}_{\rm 1,M} \ | \ \chi_{\rm 1,F},{\rm ECO}\right) &=& \delta\left(\chi^{\rm ECO}_{\rm 1,M}-\chi^{\rm ECO}_{\rm 1,F}\right) \ \Theta\left(\chicrit(\epsilon)-\chi^{\rm ECO}_{\rm 1, F}\right)  + \nn \\ &&
     \mathcal{N}\left(\chi^{\rm ECO}_{\rm 1,M}\ | \ \chicrit(\epsilon),\sigma\right) \ \Theta\left(\chi^{\rm ECO}_{\rm 1, F}-\chicrit(\epsilon)\right), 
\end{eqnarray}
where $\Theta(\cdot)$ is the Heaviside step function, $\mathcal{N}$ is a Gaussian distribution, and $\sigma$ is its standard deviation. Inserting the above equations into Eqs.~(\ref{eq:mergerspin1})-(\ref{eq:mergerspin}), we obtain the spin distribution at merger for ECOs, namely
\begin{eqnarray}
    p_{\rm pop}\left(\chi^{\rm ECO}_{\rm 1,M } \ | \ \alpha,\beta, {\rm ECO}\right) &=& \left(1-\lambda_{\rm ECO}\right) \ \beta_{\rm F}\left(\chi^{\rm ECO}_{\rm 1,M} \ | \ \alpha,\beta\right)+ \nn \\ &&\lambda_{\rm ECO} \ \mathcal{N} \left(\chi^{\rm ECO}_{\rm 1,M} \ | \ \chicrit(\epsilon),\sigma\right) \,,
    \label{eq:spinfinal}
\end{eqnarray}
where we have defined
\begin{equation}
    \lambda_{\rm ECO} (\epsilon) \coloneqq \int_{\chicrit(\epsilon)}^1 \de \chi^{\rm ECO}_{\rm 1,F} \,  \beta_{\rm F}\left(\chi^{\rm ECO}_{\rm 1,F} \ | \ \alpha,\beta\right)
\end{equation}
as the fraction of ECOs subjected to the ergoregion instability.

\begin{figure}[t]
    \centering
    \includegraphics{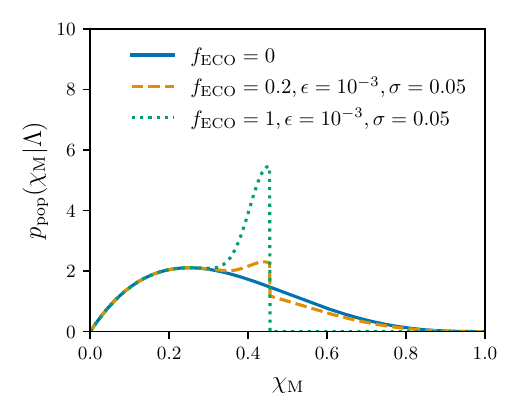}
    \caption{Spin magnitude distribution for one of the two compact objects in the binary with difference choices of the fraction of ECOs composing the population, the ECO compactness $\epsilon$, and the parameter $\sigma$ governing the agnostic gaussian distribution for the spin-down process of the ergoregion insability. On the vertical axis, $\Lambda$ indicates a common set of population parameters.}
    \label{fig:spinex}
\end{figure}

Fig.~\ref{fig:spinex} shows the spin magnitude distribution for one of the two objects in a binary, $p_{\rm pop}(\chi_{\rm 1,M}|\alpha,\beta)$, for different assumptions on the population parameters. If the fraction of ECOs composing the population is null, the spin magnitude distribution will be equal to the beta function at formation. Differently, if ECOs compose the totality of the population, and have a compactness $\epsilon=10^{-3}$ as an example, we will observe a spin magnitude distribution with an overdensity of compact objects around the value of $\chicrit \sim 0.42$. Finally, the population can be a mixture of the two cases.
    
\subsection{Hierarchical Bayesian inference and rate model} \label{sec:stat_model}

The problem of reconstructing the population properties from a set of GW observations is equivalent to finding the CBC merger rate model
\begin{equation}
    \frac{\de \Ncbc}{\de \theta \ \de z \ \de t_s} \left(\Lambda\right),
    \label{eq:merger_rate}
\end{equation}
which describes the detected sources~\cite{Vitale:2020aaz}. The CBC merger rate is a central quantity to the inference problem and is a function of a set of population parameters $\Lambda=\{ \vec{\lambda}, R_0 \}$, where $R_0$ is the source production rate today, and $\vec{\lambda}$ includes a common set of population parameters to describe the rate, such as $\epsilon$. The CBC rate describes the number density of CBC mergers in a source time $t_s$ per redshift $z$ and binary parameters $\theta$. We parametrize the CBC merger rate as 
\begin{equation}
     \frac{\de \Ncbc}{\de \theta \ \de z \ \de t_s} (\Lambda) = R_0 \, p_{\rm pop}(\theta, z |  \vec{\lambda} ) ,
\end{equation}
where $p_{\rm pop}(\theta, z | \vec{\lambda} )$ is a fiducial population model parameterized by $\vec{\lambda}$, and represents the probability density that a component of the population has binary parameters $\theta$ and is at redshift $z$. The main goal is the inference of the population parameters, which can be tackled in the framework of hierarchical Bayesian inference \cite{Mandel:2018mve, Vitale:2020aaz}.
In this work, the binary parameters of interest for the population fit are the two source masses \mos, \mts and the two spin magnitudes at merger \com and \ctm. The masses, spins and redshift distributions are specified by a separate set of population parameters: $\Lambda_r$ for the redshift, $\Lambda_m$ for the source masses and $\Lambda_s$ for the spins. The rate parametrization is
\begin{eqnarray}
    \frac{\de \Ncbc}{\de \mos \de \mts \de \com \de \ctm \de z \de t_s} (\Lambda) &=& R_0 \psi(z;\Lambda_r) \frac{\de V_c}{\de z}  p_{\rm pop}(\mos,\mts|\Lambda_m) \times \nn \\ && p_{\rm pop}(\com|\Lambda_s)p_{\rm pop}(\ctm|\Lambda_s) \,,
    \label{eq:allrate}
\end{eqnarray}
where $R_0$ is the CBC merger rate per year per comoving volume, $\psi(z;\Lambda_r)$ is a function parametrizing the CBC rate evolution in redshift, $p_{\rm pop}(\mos,\mts|\Lambda_m)$ is the spectrum of the primary and secondary masses, $p_{\rm pop}(\com|\Lambda_s)$ and $p_{\rm pop}(\ctm|\Lambda_s)$ are the distributions of the spin magnitudes for the primary and secondary objects. 

For the redshift and mass distributions, we assume the phenomenological models that are typically used for population analyses. We use a \textsc{Power Law + Peak} \cite{Talbot:2018cva} with parameters consistent with the latest studies of Ref.~\cite{KAGRA:2021duu}. The CBC merger rate evolution in redshift is parametrized using an analytical form similar to the Star formation rate, see~\ref{app:A} for more details. We further assume a flat $\Lambda$CDM cosmology with $H_0=67.7$ km/s/Mpc and $\Omega_m=0.308$ \cite{Planck:2015fie}. More details about the mass and redshift distributions are given in~\ref{app:A}. For the spin distribution, we use the model in Eq.~(\ref{eq:ffff}) that we constructed in Sec.~\ref{sec:2} for a mixture population of BHs and ECOs.
It is important to notice that, our rate model factorizes the redshift, spin magnitude and mass distributions as independent from each other. The direct consequence of this assumption is that we assume BHs and ECOs to have the same mass and redshift distribution. This could be the case if ECOs and BHs are formed via similar astrophysical channels (see \cite{2021hgwa.bookE..16M} for a review). As there is no observational indication on the astrophysical formation channels for ECOs, we make this simplistic assumption.   

To understand whether a rate model can describe the population of GW events, we need a likelihood model. The process detecting \Ngw events from the data \xdata can be described as an inhomogeneous Poissonian process \cite{Mandel:2018mve, Vitale:2020aaz} with likelihood 
\begin{eqnarray}
    \mathcal{L}\left(\{x\}|\Lambda\right) &\propto&  e^{-N_{\rm exp}(\Lambda)} \prod_i^{\Ngw} \Tobs \int \de \theta  \de z \; \mathcal{L}_{\rm GW}(x_i|\theta,z)  \ \frac{1}{1+z} \ \frac{\de \Ncbc}{\de \theta \ \de z \ \de t_s}(\Lambda),
    \label{eq:fund}
\end{eqnarray}
where $T_{\rm obs}$ is the observing time and $\mathcal{L}_{\rm GW}(x_i|\theta,z)$ is the GW likelihood, which is linked to the noise model of the detectors. The last factor can be, for instance, the rate parametrization in Eq.~\ref{eq:allrate}. In the above equation, \Nexp is the expected number of detected events, and it accounts for the presence of selection biases. Given this so-called hyperlikelihood and a prior distribution for the hyperparameters $\Lambda$, we are allowed to compute the corresponding hyperposterior $p(\Lambda|\{x\})$ through the Bayes theorem. We delegated to the Python package \icarogw~\cite{Mastrogiovanni:2023zbw} the reconstruction of this posterior distribution, from which we gain information on the parameters $\Lambda$, and in turn on the population of merging binaries. Some technical and numerical details about these computations are collected in~\ref{app:B}.

\section{Constraints from the LIGO, Virgo, KAGRA third Observing run}
\label{sec:4}

We exploit the latest CBC sources reported during the third observing run, O3, to inspect for non-trivial structures due to the ECOs' presence in the spin distribution of the population of compact objects.

\subsection{Data and common statistical tools}

We consider all the confident CBC detections with a False Alarm rate (FAR) lower than $0.25 \ \rm{yr^{-1}}$ from the third observing run (O3) of the GW detectors \cite{2019PhRvL.123w1107T,2019PhRvL.123w1108A,2020PhRvD.102f2003B,Virgo:2022fxr}.  The choice corresponds to a selection of 52 binaries. To compute the hierarchical likelihood, we use the parameter estimation samples~\footnote{We use parameter estimation samples that are obtained by mixing samples from the \texttt{IMRPhenom} \cite{Thompson:2020nei,Pratten:2020ceb} and \texttt{SEOBNR} \cite{Ossokine:2020kjp,Matas:2020wab} family waveforms.}  on the two compact objects' spin magnitudes, source masses and redshift released with the GWTC-2.1 \cite{LIGOScientific:2021usb} for the events observed in the first half of O3 and with the GWTC-3 for the second half of O3 \cite{KAGRA:2021vkt}.
The spin magnitude at the merger of each compact object in the GW events is reported in the first two columns in Tab.~\ref{tab:tab_spin}. The other columns report the probability for the object to be an ECO after performing our analysis. For most of the events detected during the third LVK observing run, it is impossible to constrain the spin magnitudes of the two objects significantly. However, for a few cases, such as GW190517\_055101 and GW191109\_010717, it is possible to measure a high spin magnitude.

\begin{table}
\centering
{\scriptsize
\begin{tabular}[t]{||l|c|c|c|c|c|c||}
    \hline
    GW event & $\chi_{\rm 1, M}$ & $\chi_{\rm 2, M}$ & $p_{1,p}({\rm ECO}|\{x\})$ & $p_{2,p}({\rm ECO}|\{x\})$ & $p_{1,a}({\rm ECO}|\{x\})$ & $p_{2,a}({\rm ECO}|\{x\})$ \\ 
    \hline
    \hline
    GW190408\_181802 & ${0.31}^{+0.51}_{-0.28}$ & ${0.37}^{+0.53}_{-0.33}$& $0.42^{+0.44}_{-0.37}$ & $0.38^{+0.45}_{-0.34}$ & $0.33^{+0.42}_{-0.28}$ & $0.30^{+0.42}_{-0.26}$ \\ 
GW190412\_053044 & ${0.31}^{+0.22}_{-0.19}$ & ${0.40}^{+0.50}_{-0.37}$& $0.15^{+0.77}_{-0.14}$ & $0.34^{+0.51}_{-0.30}$ & $0.12^{+0.62}_{-0.11}$ & $0.28^{+0.43}_{-0.24}$ \\ 
GW190413\_134308 & ${0.55}^{+0.40}_{-0.49}$ & ${0.49}^{+0.45}_{-0.44}$& $0.32^{+0.43}_{-0.29}$ & $0.33^{+0.44}_{-0.30}$ & $0.26^{+0.39}_{-0.23}$ & $0.26^{+0.40}_{-0.23}$ \\ 
GW190421\_213856 & ${0.42}^{+0.50}_{-0.38}$ & ${0.46}^{+0.47}_{-0.41}$& $0.35^{+0.45}_{-0.32}$ & $0.35^{+0.44}_{-0.31}$ & $0.28^{+0.41}_{-0.24}$ & $0.28^{+0.40}_{-0.24}$ \\ 
GW190503\_185404 & ${0.42}^{+0.48}_{-0.38}$ & ${0.44}^{+0.49}_{-0.39}$& $0.36^{+0.45}_{-0.32}$ & $0.36^{+0.45}_{-0.32}$ & $0.28^{+0.41}_{-0.25}$ & $0.28^{+0.41}_{-0.25}$ \\ 
GW190512\_180714 & ${0.21}^{+0.50}_{-0.19}$ & ${0.40}^{+0.51}_{-0.37}$& $0.49^{+0.43}_{-0.44}$ & $0.37^{+0.44}_{-0.33}$ & $0.38^{+0.45}_{-0.33}$ & $0.29^{+0.42}_{-0.25}$ \\ 
GW190513\_205428 & ${0.44}^{+0.45}_{-0.40}$ & ${0.46}^{+0.47}_{-0.42}$& $0.32^{+0.45}_{-0.29}$ & $0.33^{+0.45}_{-0.29}$ & $0.25^{+0.40}_{-0.22}$ & $0.27^{+0.40}_{-0.23}$ \\ 
\textbf{GW190517\_055101} & ${0.89}^{+0.09}_{-0.31}$ & ${0.65}^{+0.32}_{-0.56}$& $0.00^{+0.04}_{-0.00}$ & $0.18^{+0.42}_{-0.17}$ & $0.00^{+0.10}_{-0.00}$ & $0.16^{+0.35}_{-0.14}$ \\ 
GW190519\_153544 & ${0.65}^{+0.30}_{-0.47}$ & ${0.58}^{+0.37}_{-0.52}$& $0.08^{+0.36}_{-0.08}$ & $0.25^{+0.39}_{-0.23}$ & $0.07^{+0.34}_{-0.06}$ & $0.20^{+0.35}_{-0.18}$ \\ 
GW190521\_030229 & ${0.71}^{+0.27}_{-0.61}$ & ${0.50}^{+0.44}_{-0.45}$& $0.29^{+0.37}_{-0.26}$ & $0.30^{+0.44}_{-0.26}$ & $0.25^{+0.34}_{-0.22}$ & $0.25^{+0.38}_{-0.21}$ \\ 
GW190521\_074359 & ${0.37}^{+0.46}_{-0.33}$ & ${0.44}^{+0.47}_{-0.39}$& $0.40^{+0.44}_{-0.36}$ & $0.34^{+0.46}_{-0.31}$ & $0.31^{+0.42}_{-0.27}$ & $0.27^{+0.42}_{-0.23}$ \\ 
GW190527\_092055 & ${0.41}^{+0.51}_{-0.37}$ & ${0.40}^{+0.51}_{-0.36}$& $0.35^{+0.47}_{-0.32}$ & $0.36^{+0.46}_{-0.33}$ & $0.28^{+0.41}_{-0.24}$ & $0.28^{+0.40}_{-0.25}$ \\ 
GW190602\_175927 & ${0.50}^{+0.44}_{-0.45}$ & ${0.53}^{+0.42}_{-0.47}$& $0.31^{+0.44}_{-0.28}$ & $0.32^{+0.42}_{-0.29}$ & $0.25^{+0.40}_{-0.22}$ & $0.26^{+0.38}_{-0.22}$ \\ 
GW190620\_030421 & ${0.71}^{+0.25}_{-0.53}$ & ${0.57}^{+0.38}_{-0.51}$& $0.11^{+0.32}_{-0.10}$ & $0.28^{+0.39}_{-0.25}$ & $0.10^{+0.33}_{-0.09}$ & $0.23^{+0.36}_{-0.20}$ \\ 
GW190630\_185205 & ${0.27}^{+0.45}_{-0.25}$ & ${0.44}^{+0.47}_{-0.39}$& $0.38^{+0.49}_{-0.34}$ & $0.33^{+0.46}_{-0.29}$ & $0.30^{+0.45}_{-0.26}$ & $0.26^{+0.40}_{-0.23}$ \\ 
GW190701\_203306 & ${0.44}^{+0.48}_{-0.40}$ & ${0.45}^{+0.48}_{-0.41}$& $0.36^{+0.44}_{-0.32}$ & $0.35^{+0.45}_{-0.32}$ & $0.28^{+0.41}_{-0.25}$ & $0.28^{+0.41}_{-0.24}$ \\ 
GW190706\_222641 & ${0.68}^{+0.28}_{-0.53}$ & ${0.54}^{+0.41}_{-0.48}$& $0.17^{+0.38}_{-0.15}$ & $0.29^{+0.42}_{-0.26}$ & $0.14^{+0.36}_{-0.12}$ & $0.24^{+0.40}_{-0.21}$ \\ 
GW190707\_093326 & ${0.20}^{+0.47}_{-0.18}$ & ${0.33}^{+0.54}_{-0.30}$& $0.49^{+0.43}_{-0.44}$ & $0.40^{+0.45}_{-0.36}$ & $0.38^{+0.45}_{-0.32}$ & $0.30^{+0.44}_{-0.27}$ \\ 
GW190708\_232457 & ${0.23}^{+0.48}_{-0.21}$ & ${0.34}^{+0.53}_{-0.31}$& $0.45^{+0.46}_{-0.40}$ & $0.40^{+0.45}_{-0.36}$ & $0.34^{+0.45}_{-0.30}$ & $0.31^{+0.43}_{-0.27}$ \\ 
GW190720\_000836 & ${0.36}^{+0.37}_{-0.31}$ & ${0.47}^{+0.45}_{-0.42}$& $0.20^{+0.58}_{-0.18}$ & $0.28^{+0.43}_{-0.25}$ & $0.16^{+0.46}_{-0.14}$ & $0.23^{+0.37}_{-0.20}$ \\ 
GW190727\_060333 & ${0.52}^{+0.42}_{-0.47}$ & ${0.47}^{+0.47}_{-0.42}$& $0.32^{+0.42}_{-0.29}$ & $0.34^{+0.43}_{-0.31}$ & $0.25^{+0.39}_{-0.22}$ & $0.27^{+0.40}_{-0.24}$ \\ 
GW190728\_064510 & ${0.33}^{+0.38}_{-0.29}$ & ${0.37}^{+0.51}_{-0.34}$& $0.25^{+0.59}_{-0.23}$ & $0.33^{+0.49}_{-0.30}$ & $0.21^{+0.48}_{-0.18}$ & $0.27^{+0.42}_{-0.23}$ \\ 
GW190803\_022701 & ${0.41}^{+0.50}_{-0.37}$ & ${0.44}^{+0.49}_{-0.40}$& $0.37^{+0.45}_{-0.33}$ & $0.36^{+0.44}_{-0.32}$ & $0.29^{+0.42}_{-0.25}$ & $0.28^{+0.41}_{-0.25}$ \\ 
GW190828\_063405 & ${0.53}^{+0.40}_{-0.46}$ & ${0.46}^{+0.46}_{-0.41}$& $0.25^{+0.45}_{-0.23}$ & $0.31^{+0.45}_{-0.28}$ & $0.21^{+0.38}_{-0.18}$ & $0.25^{+0.39}_{-0.22}$ \\ 
GW190828\_065509 & ${0.27}^{+0.45}_{-0.24}$ & ${0.44}^{+0.49}_{-0.40}$& $0.40^{+0.50}_{-0.36}$ & $0.35^{+0.45}_{-0.32}$ & $0.31^{+0.46}_{-0.27}$ & $0.28^{+0.41}_{-0.25}$ \\ 
GW190910\_112807 & ${0.34}^{+0.52}_{-0.31}$ & ${0.37}^{+0.53}_{-0.34}$& $0.40^{+0.44}_{-0.36}$ & $0.38^{+0.46}_{-0.34}$ & $0.31^{+0.43}_{-0.27}$ & $0.30^{+0.42}_{-0.26}$ \\ 
GW190915\_235702 & ${0.55}^{+0.39}_{-0.49}$ & ${0.45}^{+0.48}_{-0.41}$& $0.31^{+0.42}_{-0.28}$ & $0.35^{+0.44}_{-0.31}$ & $0.25^{+0.39}_{-0.22}$ & $0.28^{+0.40}_{-0.24}$ \\ 
GW190924\_021846 & ${0.23}^{+0.46}_{-0.21}$ & ${0.33}^{+0.53}_{-0.30}$& $0.49^{+0.42}_{-0.44}$ & $0.42^{+0.44}_{-0.39}$ & $0.37^{+0.44}_{-0.32}$ & $0.32^{+0.44}_{-0.28}$ \\ 
GW190925\_232845 & ${0.36}^{+0.50}_{-0.32}$ & ${0.41}^{+0.50}_{-0.37}$& $0.36^{+0.48}_{-0.33}$ & $0.36^{+0.46}_{-0.33}$ & $0.29^{+0.43}_{-0.25}$ & $0.29^{+0.42}_{-0.25}$ \\ 
GW190929\_012149 & ${0.34}^{+0.54}_{-0.31}$ & ${0.47}^{+0.46}_{-0.43}$& $0.39^{+0.45}_{-0.35}$ & $0.35^{+0.43}_{-0.31}$ & $0.30^{+0.43}_{-0.26}$ & $0.27^{+0.40}_{-0.24}$ \\ 
GW190930\_133541 & ${0.40}^{+0.38}_{-0.35}$ & ${0.46}^{+0.46}_{-0.41}$& $0.22^{+0.55}_{-0.20}$ & $0.32^{+0.44}_{-0.29}$ & $0.19^{+0.44}_{-0.16}$ & $0.26^{+0.38}_{-0.22}$ \\ 
GW191105\_143521 & ${0.23}^{+0.53}_{-0.21}$ & ${0.34}^{+0.54}_{-0.31}$& $0.49^{+0.42}_{-0.43}$ & $0.41^{+0.44}_{-0.37}$ & $0.37^{+0.44}_{-0.32}$ & $0.32^{+0.43}_{-0.28}$ \\ 
\textbf{GW191109\_010717} & ${0.82}^{+0.16}_{-0.58}$ & ${0.64}^{+0.33}_{-0.57}$& $0.14^{+0.35}_{-0.12}$ & $0.28^{+0.34}_{-0.25}$ & $0.12^{+0.33}_{-0.10}$ & $0.23^{+0.33}_{-0.20}$ \\ 
GW191127\_050227 & ${0.67}^{+0.30}_{-0.57}$ & ${0.55}^{+0.41}_{-0.49}$& $0.24^{+0.41}_{-0.21}$ & $0.31^{+0.42}_{-0.28}$ & $0.20^{+0.35}_{-0.17}$ & $0.25^{+0.38}_{-0.22}$ \\ 
GW191129\_134029 & ${0.25}^{+0.37}_{-0.22}$ & ${0.35}^{+0.51}_{-0.31}$& $0.41^{+0.50}_{-0.37}$ & $0.38^{+0.47}_{-0.34}$ & $0.32^{+0.47}_{-0.28}$ & $0.30^{+0.43}_{-0.26}$ \\ 
GW191204\_171526 & ${0.40}^{+0.38}_{-0.35}$ & ${0.46}^{+0.41}_{-0.40}$& $0.17^{+0.57}_{-0.16}$ & $0.27^{+0.46}_{-0.24}$ & $0.14^{+0.46}_{-0.13}$ & $0.22^{+0.39}_{-0.19}$ \\ 
GW191215\_223052 & ${0.47}^{+0.45}_{-0.42}$ & ${0.44}^{+0.49}_{-0.40}$& $0.35^{+0.43}_{-0.32}$ & $0.33^{+0.46}_{-0.30}$ & $0.28^{+0.40}_{-0.24}$ & $0.27^{+0.42}_{-0.24}$ \\ 
GW191216\_213338 & ${0.24}^{+0.36}_{-0.22}$ & ${0.36}^{+0.50}_{-0.32}$& $0.36^{+0.56}_{-0.33}$ & $0.31^{+0.54}_{-0.27}$ & $0.28^{+0.49}_{-0.25}$ & $0.26^{+0.44}_{-0.22}$ \\ 
GW191222\_033537 & ${0.38}^{+0.51}_{-0.35}$ & ${0.42}^{+0.50}_{-0.39}$& $0.39^{+0.44}_{-0.35}$ & $0.37^{+0.43}_{-0.34}$ & $0.30^{+0.42}_{-0.26}$ & $0.29^{+0.42}_{-0.25}$ \\ 
GW191230\_180458 & ${0.51}^{+0.44}_{-0.46}$ & ${0.49}^{+0.45}_{-0.44}$& $0.32^{+0.46}_{-0.29}$ & $0.33^{+0.45}_{-0.29}$ & $0.26^{+0.41}_{-0.23}$ & $0.26^{+0.41}_{-0.23}$ \\ 
GW200112\_155838 & ${0.35}^{+0.47}_{-0.31}$ & ${0.39}^{+0.51}_{-0.35}$& $0.36^{+0.47}_{-0.33}$ & $0.36^{+0.46}_{-0.33}$ & $0.29^{+0.42}_{-0.25}$ & $0.28^{+0.42}_{-0.25}$ \\ 
GW200128\_022011 & ${0.60}^{+0.36}_{-0.53}$ & ${0.50}^{+0.44}_{-0.45}$& $0.30^{+0.40}_{-0.27}$ & $0.33^{+0.41}_{-0.30}$ & $0.24^{+0.38}_{-0.21}$ & $0.26^{+0.39}_{-0.23}$ \\ 
GW200129\_065458 & ${0.54}^{+0.41}_{-0.48}$ & ${0.49}^{+0.43}_{-0.43}$& $0.32^{+0.45}_{-0.28}$ & $0.25^{+0.52}_{-0.22}$ & $0.25^{+0.39}_{-0.22}$ & $0.20^{+0.42}_{-0.18}$ \\ 
GW200202\_154313 & ${0.22}^{+0.45}_{-0.20}$ & ${0.33}^{+0.53}_{-0.30}$& $0.47^{+0.45}_{-0.42}$ & $0.41^{+0.45}_{-0.37}$ & $0.36^{+0.45}_{-0.31}$ & $0.32^{+0.44}_{-0.28}$ \\ 
GW200208\_130117 & ${0.36}^{+0.51}_{-0.32}$ & ${0.42}^{+0.49}_{-0.38}$& $0.37^{+0.47}_{-0.33}$ & $0.35^{+0.46}_{-0.31}$ & $0.29^{+0.42}_{-0.26}$ & $0.28^{+0.41}_{-0.24}$ \\ 
GW200209\_085452 & ${0.52}^{+0.43}_{-0.46}$ & ${0.49}^{+0.45}_{-0.44}$& $0.32^{+0.43}_{-0.29}$ & $0.32^{+0.43}_{-0.29}$ & $0.26^{+0.39}_{-0.23}$ & $0.26^{+0.40}_{-0.23}$ \\ 
GW200219\_094415 & ${0.48}^{+0.45}_{-0.43}$ & ${0.47}^{+0.47}_{-0.42}$& $0.33^{+0.44}_{-0.30}$ & $0.33^{+0.45}_{-0.30}$ & $0.27^{+0.40}_{-0.23}$ & $0.27^{+0.40}_{-0.23}$ \\ 
GW200224\_222234 & ${0.47}^{+0.44}_{-0.42}$ & ${0.44}^{+0.48}_{-0.39}$& $0.30^{+0.46}_{-0.27}$ & $0.33^{+0.46}_{-0.30}$ & $0.24^{+0.41}_{-0.21}$ & $0.26^{+0.41}_{-0.23}$ \\ 
GW200225\_060421 & ${0.58}^{+0.35}_{-0.51}$ & ${0.43}^{+0.49}_{-0.39}$& $0.27^{+0.40}_{-0.24}$ & $0.34^{+0.44}_{-0.30}$ & $0.22^{+0.37}_{-0.19}$ & $0.28^{+0.40}_{-0.24}$ \\ 
GW200302\_015811 & ${0.38}^{+0.51}_{-0.35}$ & ${0.45}^{+0.48}_{-0.41}$& $0.37^{+0.47}_{-0.33}$ & $0.34^{+0.46}_{-0.31}$ & $0.29^{+0.42}_{-0.25}$ & $0.27^{+0.42}_{-0.24}$ \\ 
GW200311\_115853 & ${0.40}^{+0.48}_{-0.36}$ & ${0.41}^{+0.50}_{-0.37}$& $0.36^{+0.46}_{-0.33}$ & $0.36^{+0.46}_{-0.33}$ & $0.29^{+0.42}_{-0.25}$ & $0.29^{+0.42}_{-0.25}$ \\ 
GW200316\_215756 & ${0.33}^{+0.37}_{-0.29}$ & ${0.43}^{+0.47}_{-0.38}$& $0.30^{+0.55}_{-0.27}$ & $0.31^{+0.48}_{-0.28}$ & $0.24^{+0.46}_{-0.21}$ & $0.25^{+0.41}_{-0.22}$ 
\\
\hline 
\hline 
\end{tabular}
    \caption{The table reports the median and 90\% symmetric C.I. for the spin magnitudes of the two compact objects (second and third columns), the probability for the objects to be ECOs under polar and axial perturbations (fourth to seventh columns) for all the GW events detected during O3.}
    \label{tab:tab_spin}
}
\end{table}

We study three different spin population models described in Eq.~(\ref{eq:spinfinal}): the \bbh model in which the CBC population is composed only of classical BHs, the \eco model in which the CBC population is composed only of ECOs and the \mixture model in which the CBC population is a mixture of BHs and ECOs.  
The spin population model has 6 hyper-parameters: the two $\beta$-distribution parameters at formation, $\alpha$ and $\beta$, the fraction of ECOs in the population, $f_{\rm ECO}$, the ECO's compactness, $\epsilon$, and the width of the gaussian distribution around the critical spin for the ergoregion instability, $\sigma$. In Table~\ref{tab:priors}, we report the priors used on the population parameters for the three models.
\begin{table}
\centering
{\small
\begin{tabular}[t]{||c c c c||}
 \hline
 \textbf{Parameter} & \bbh & \eco & \mixture \\ [0.5ex] 
 \hline\hline
 $\alpha_{\rm F}$ & $\mathcal{U}$[1.1,10] & $\mathcal{U}$[1.1,10] & $\mathcal{U}$[1.1,10] \\ 
 \hline
$\beta_{\rm F}$ & $\mathcal{U}$[1.1,10] & $\mathcal{U}$[1.1,10] & $\mathcal{U}$[1.1,10] \\
 \hline
 \feco & 0 & 1 & $\mathcal{U}$[0,1] \\
 \hline
 $\epsilon$ & - & $\mathcal{LU}[10^{-42},10^{-3}]$ & $\mathcal{LU}[10^{-42},10^{-3}]$ \\
 \hline
 $\sigma$ & - & $\mathcal{U}$[0.005,0.5] & $\mathcal{U}$[0.005,0.5] \\ [1ex] 
 \hline
 \end{tabular}
 \caption{Priors used on the population parameters describing the spin distribution in Eq.~(\ref{eq:ffff}) for the three population models. Uniform priors are indicated with $\mathcal{U}$ and logarithmic uniform priors with $\mathcal{LU}$. The prior range on $\epsilon$ can be mapped into the prior boundary on \chicrit following Eq.~(\ref{eq:chicrit}). For polar perturbations, \chicrit $\in [0.03 ,0.45]$ while for axial perturbations \chicrit $\in [0.05, 0.68]$.}
\label{tab:priors}
}
\end{table}
We verified that fitting also for other population parameters related to the redshift and mass distribution does not significantly change the results obtained on the spin magnitude distributions. For the runs, we fix the \textsc{Power Law + peak} parameters to $\alpha_M=3.0$, $\beta_M=0.8$, $m_{\rm min}=4.5 M_\odot$, $m_{\rm max}=100 M_\odot$, $\delta_m=4.8 M_\odot$, $\mu_g=35.5 M_\odot$, $\sigma_g=4.8 M_\odot$ and $\lambda_{\rm peak}=0.05$ and the merger rate redshift parameters as $\gamma=2.7$, $k=3$, $z_p=2.0$. We refer the reader to~\ref{app:A} for the definitions of the mass and redshift population parameters.

We use the Bayes factors to discuss the preference among competing models (hypotheses). The Bayes factor is defined as 
\begin{equation}
    \mathcal{B}=\frac{\int \mathcal{L}(\{x\}|\Lambda,\mathcal{H}_1) \ p(\Lambda|,\mathcal{H}_1) \ \de \Lambda}{\int \mathcal{L}(\{x\}|\Lambda,\mathcal{H}_2) \ p(\Lambda|\mathcal{H}_2) \ \de \Lambda},
\end{equation}
where $\mathcal{H}_1$ and $\mathcal{H}_2$ are two competing models (hypotheses),  $\mathcal{L}(\{x\}|\Lambda,\mathcal{H}_2)$ is the hierarchical likelihood in Eq.~(\ref{eq:fund}) calculated using the population model for the hypothesis $\mathcal{H}_2$ and $p(\Lambda|\mathcal{H}_2)$ is a prior on the population parameters. We also use posterior predictive checks to compare the reconstructed spin distributions by each model. We define the reconstructed spin distribution as 
\begin{equation}
    p_{\rm pop}(\chi_{\rm M}|\{x\})= \int p_{\rm pop}(\chi_{\rm M}|\Lambda) p(\Lambda|\{x\}) \de \Lambda, 
\end{equation}
where $p(\Lambda|\{x\})$ is the hierarchical posterior on a model's population parameters. Moreover, for each compact object, $y$, we define the probability to be an ECO given the population fit from data $\{x\}$ as
\begin{eqnarray}
    &&p({\rm ECO }|y, \{x\})=\int p(\Lambda|\{x\})  p({\rm ECO}|\chi_{\rm M},\Lambda) p(\chi_{\rm M}|y,\Lambda) \de \chi_{\rm M}  d\Lambda \nn \\
    &&=\int p(\Lambda|\{x\})  \frac{p(\chi_{\rm M}|{\rm ECO},\Lambda)p({\rm ECO}|\Lambda)}{p(\chi_{\rm M}|{\rm ECO},\Lambda)+p(\chi_{\rm M}|{\rm BH},\Lambda)} p(\chi_{\rm M}|y,\Lambda)  \de \chi_{\rm M} \de \Lambda. \nn \\
    \label{eq:ECOprob}
\end{eqnarray}
In the above equation, $p(\chi_{\rm M}|{\rm ECO})$ and $p(\chi_{\rm M}|{\rm BH})$ are the ECO and BH spin magnitude distributions defined in Eqs.~\ref{eq:spinfinal} and~(\ref{eq:spinfinal1}), respectively. The factor $p({\rm ECO}|\Lambda)$ is the probability that an event is generated as an ECO, i.e.,  $p({\rm ECO}|\Lambda)=f_{\rm ECO}$. The factor $p(\chi_{\rm M}|y,\Lambda)$  is the posterior probability on $\chi_{\rm M}$ for the event $y$  given the population model with parameters $\Lambda$, and $p(\Lambda|\{x\})$ is the hierarchical posterior. Note that, for Eq.~(\ref{eq:ECOprob}) to be formally correct, the event $y$  should not be included in the set $\{x\}$ from which the hierarchical posterior is computed. However, here we neglect this assumption as the number of compact objects is high enough that the inclusion of $y$  in the fitting data does not change significantly the population results.

\subsection{Results from the LVK third observing run}

In Tab.~\ref{tab:bf}, we provide  the Bayes factors between several population models and the \bbh one. 
\begin{table}
\centering
{
\small
\begin{tabular}[t]{||c c||}
 \hline
 \bbh \: \textit{vs} & $\log_{10}\mathcal{B}$ \\
 \hline\hline
 \eco (polar) & $+\infty$ \\ 
 \eco (axial) & $-0.08$ \\
 \mixture (axial) & $-0.24$ \\
 \mixture (polar) & $-0.41$ \\
 \hline
 \end{tabular}
 \caption{Bayes factors for the BBH population model versus the ECO and ECO-BH mixture population models.}
\label{tab:bf}
}
\end{table}
Given the prior ranges in Tab.~\ref{tab:priors} for the parameter $\epsilon$, we find that the only model that we are able to exclude is the \eco model in the case of polar perturbations. This result is due to the fact that the maximum spin that the \eco polar population predicts is $\chi_{\rm{crit}} = 0.45$ (see Fig.~\ref{fig:chicrit}), which is inconsistent with the  spin of the primary object in GW190517\_055101 (see Tab.~\ref{tab:tab_spin}). Instead, we cannot exclude the \eco axial model for which the maximum spin is $\chi_{\rm{crit}}=0.68$, which is consistent with the spin measurement of GW190517\_055101. With the current data, we cannot also discriminate between the \bbh model and any of the \mixture models. The fact that we cannot rule out the \eco (axial) and \mixture models with respect to the simple \bbh model is indicative of the fact that there is no support of peak-like structures in the spin distribution of the CBCs with current GW detections~\cite{Kimball:2020opk, Callister:2022qwb,Tong:2022iws, Mould:2022xeu,Adamcewicz:2023szp}. Even though we cannot rule out some of the models, we can still draw interesting population-driven constraints on the ECOs properties.

\begin{figure}[t]
    \centering
    \includegraphics{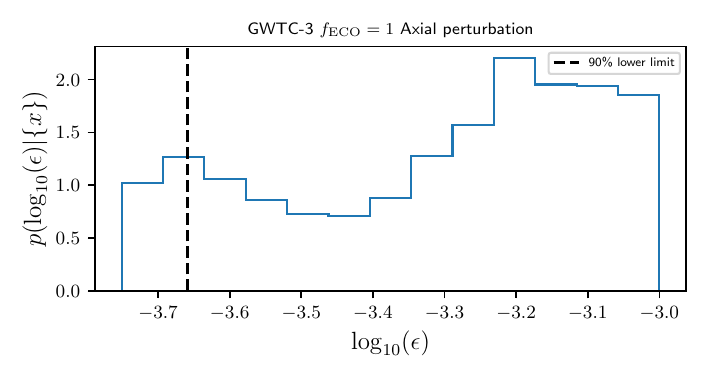}
    \caption{Marginal posterior on $\log_{10}(\epsilon)$ for the case in which the population of compact objects observed during O3 is composed by ECOs with totally reflective surfaces. The dashed vertical line indicates the 90\% credible level lower limit.}
    \label{fig:totally_reflective}
\end{figure}
Fig.~\ref{fig:totally_reflective} shows the posterior on $\epsilon$ that we obtain in the case in which the population of CBCs is composed of 100 \% of ECOs for axial perturbations. In this case, we obtain it is possible to set a lower limit on $\epsilon$, which is $\log_{10}(\epsilon)>-3.65$ at 90\% C.I. In other words, when we assume that the entire population of LVK detections is composed of totally reflective ECOs, we must rule out that these objects are ultra-compact. This type of constraint is mostly driven by the fact that the primary objects of GW190517\_055101 and GW191109\_010717 are fastly spinning.

\begin{figure}[t]
    \centering
    \includegraphics{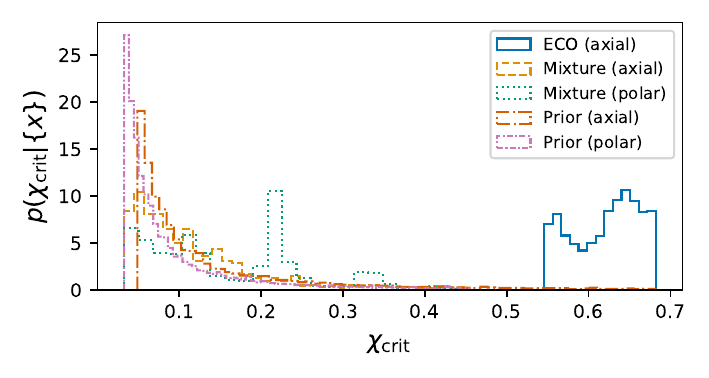}
    \caption{Implied posteriors from the events detected during O3 on the critical spin magnitude \chicrit at which the ergoregion instability is expected to happen. The color lines indicate the different population models.}
    \label{fig:totally_reflective_chi_crit}
\end{figure}
The interplay between $\log_{10}(\epsilon)$ and the implied $\chi_{\rm crit}$ can also be visualized in Fig.~\ref{fig:totally_reflective_chi_crit}. The implied posterior on \chicrit, in the limit that all the population is composed by ECOs, excludes values of $\chicrit<0.56$ at 90\% C.I. The motivation for this result is that if  $\chicrit>0.56$, the population spin magnitudes distribution would not be able to explain the existence of the primary objects of GW190517\_055101 and GW191109\_010717, which are fastly spinning. 

\begin{figure}[t]
    \centering
    \includegraphics[scale=0.8]{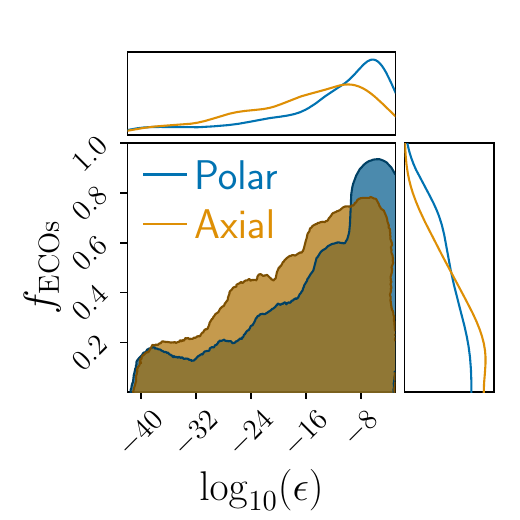}
    \includegraphics[scale=0.9]{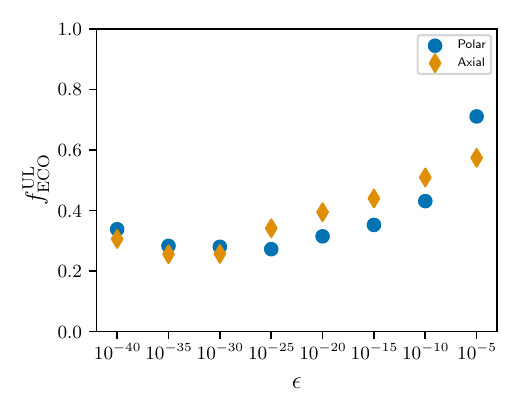}
    \caption{\textit{Left}: Joint posterior distribution on \leps and \feco from the CBCs detected during O3 with a FAR$<0.25$ yr$^{-1}$. The coloured area indicates the 90\% C.I. The figure also reports the marginal posteriors for the two population parameters. \textit{Right}: 90\% credible upper limits on the fraction of ECOs present in the CBC population inferred from the GW detections in O3. The vertical axis reports the upper limit on \feco when assuming that the population has $\epsilon$ lower than the threshold reported in the horizontal axis.}
    \label{fig:MIX_res}
\end{figure}

We now discuss the constraints on $\log_{10}(\epsilon)$ in the case that the population is composed of a mixture of BHs and ECOs. The left panel of Fig.~\ref{fig:MIX_res} displays the joint posterior on \feco and $\log_{10}(\epsilon)$. We find that it is not possible to jointly constrain the \feco and $\log_{10}(\epsilon)$ in their respective prior ranges. 
However, the joint posterior can exclude some parts of the parameter space. In particular, taking the polar perturbation case as an example, if $\feco \gtrsim 0.4$ then $\log_{10}(\epsilon) \gtrsim -16$, namely if ECOs account for 40\% of the GW signals observed, ECOs with $\log_{10}(\epsilon) \lesssim -16$ can be ruled out. Contrary, if the population of ECOs is composed of less than the 20\% of the population, then we are not able to set a meaningful constraint on \leps. The motivation for this constraint is that, if ECOs compose a small fraction of the population, highly spinning events such as GW190517\_055101 and GW191109\_010717 can be explained as BBHs. However, if the fraction of ECOs is high enough, the few observed highly spinning events cannot be explained by the population model. As we are not significantly able to set a limit on \leps for the \mixture model, the implied posterior on \chicrit is not very informative. In fact, in Fig.~\ref{fig:totally_reflective_chi_crit}, we can observe that the implied posterior on \chicrit for the \mixture models slightly deviates from the implied prior used for the analysis. 
To quantify the constraints on the possible existence of ECOs we define the 90\% upper limit (UL) on $\feco^{\rm UL}$ as
\begin{equation}
    0.9=\int_{0}^{\feco^{\rm UL}} p(\feco|\{x\}) \ \de \feco,
\end{equation}
where $p(\feco|\{x\})$ is the marginal posterior from the simulated data on $\feco$. Given the prior range on $\epsilon$ in Tab.~\ref{tab:priors}, we find that at 90\% credible interval $\feco < 71\%$ and $\feco < 59\%$ for polar and axial perturbations, respectively. 
The upper limit can also be defined as a function of $\epsilon$. In the right panel of Fig.~\ref{fig:MIX_res}, we show the upper limit on \feco computed by assuming a population of compact objects with $\epsilon$  lower than a certain threshold. If the population is of compact objects with $\epsilon<10^{-30}$ (ultra-compact objects), then no more than 28\% and 25\% of the observed population can be ECOs. The upper limits on \feco increase as we increase $\epsilon$ as the predicted $\chicrit$ increases to values consistent with the observed CBC population.

\begin{figure}
    \centering
    \includegraphics[scale=1.0]{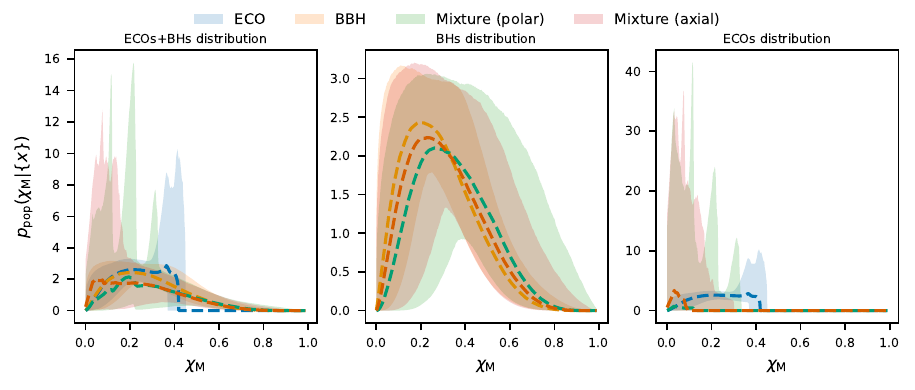}
    \caption{\textit{Left panel}: Spin magnitude distribution reconstructed for the overall population from the GW events detected during O3 for the ECOs-only, BBHs-only and mixture cases. \textit{Middle panel}: Spin magnitude distribution reconstructed for the BBH population from the GW events detected during O3 for the ECOs-only, BBHs-only and mixture cases. \textit{Right panel}: Spin magnitude distribution reconstructed for the ECOs population from the GW events detected during O3 for the ECOs-only, BBHs-only and mixture cases. The dashed lines indicate the median value of the reconstruction while the shaded area the 90\% C.I. }
    \label{fig:ppchecks_totally_reflective}
\end{figure}

Finally, let us discuss the reconstructed spin magnitude distributions. Fig.~\ref{fig:ppchecks_totally_reflective} shows the reconstructed spin magnitude distributions for all the population models. The figure also reports the contributions to the overall spin distribution given by ECOs and BHs. When we consider that the population is composed of binaries of ECOs (axial) or BHs only, the overall spin distributions that we reconstruct are similar and prefer low spin magnitudes. This result is consistent with what is obtained with the \textsc{DEFAULT} spin model in Ref.~\cite{KAGRA:2021duu}. For the BBH model, this is not surprising as effectively we are using the same spin model as in Ref.~\cite{KAGRA:2021duu}, while for the ECOs model this result is motivated by the fact that \leps is constrained to provide high values for \chicrit (see Fig.~\ref{fig:totally_reflective_chi_crit}) thus producing a spin distribution at merger very similar to the spin distribution at formation (BBH-only case) for most of the low spinning compact objects.
The average spin magnitude distribution reconstructed by the mixture model for the overall population is consistent with the one reconstructed from the ECOs and BBHs-only cases. However, the error budget of the reconstruction is higher for the low spin magnitude region. This is because the model has more degrees of freedom (\leps) that can fit the low spin magnitude region. Although the overall reconstruction of the spin distribution is consistent with the BBH-only and ECOs-only cases, there are few interesting results if we consider the population as a mixture of BHs and ECOs. According to the reconstruction of the mixture model, the BHs spin magnitude distribution slightly prefers higher spin values than the ECOs one. Instead, the ECOs spin distribution particularly prefers low values of the spin magnitudes. In other words, GW events with low spins will most likely be associated to the ECOs population, while GW events with high spins will most likely be associated to the BBH population. In Tab.~\ref{tab:tab_spin}, using the posterior predictive checks on the spin distribution, we quantify the probability that the compact objects belong to the  ECO or BBH populations of the \mixture model. We can observe from the probability that objects such as GW190517\_055101 and GW191109\_010717 have a higher probability of being classical BHs.

\section{Projections for next-generation GW detectors}
\label{sec:5}

In 2035+, we expect next-Generation GW detectors such as the Einstein Telescope \cite{2010CQGra..27s4002P, Maggiore_2020,Branchesi:2023mws} and Cosmic Explorer \cite{PhysRevD.91.082001,Reitze:2019iox} to become operative. The enhanced sensitivity of the GW detectors will provide us with thousands of GWs detected from CBCs with high signal-to-noise ratios (SNRs) per year. We study how next-Generation GW detectors can constrain the existence of ECOs from the spin distribution of the detected compact binary mergers if these are all composed of canonical BHs.

\subsection{Simulation}

We simulate two fiducial BBH populations with aligned spins ad spin magnitude at merger  described by a $\beta$ distribution with parameters $\alpha=2$, $\beta=4$ and $\alpha=6$, $\beta=2$. The former population prefers to produce BBH binaries with spin magnitudes lower than 0.5 while the latter prefers to form BBH binaries with spins higher than 0.5. We refer to these two populations as ``low spin'' and ``high spin''. The motivation for considering  populations composed of BBHs from which infer constraints on the ECOs fraction is that currently, we do not accurately know how different models of ECOs are expected to form and modify the CBC population.

We draw binaries in redshift following merger rate as a function,
\begin{equation}
    R(z)= 20 \frac{\rm mergers}{{\rm Gpc^{-3} yr^{-1}}} \left[1+(1+z_p)^{-\gamma-k}\right] \frac{(1+z)^\gamma}{1+\left(\frac{1+z}{1+z_p}\right)^{\gamma+k}}\,,
    \label{eq:rate}
\end{equation}
where for $k$ and $z_p$ we assume the fiducial values of $k=3$ and $z_p=2$, which are consistent with the Star Formation Rate \cite{2014ARAnA..52..415M}, while for $\gamma$  we take values consistent with the 90\% credible intervals values found in Ref.~\cite{KAGRA:2021duu}. The primary and secondary masses of the binaries are drawn according to a \textsc{Power Law + peak} distribution with parameters $\alpha_M=3.0$, $\beta_M=0.8$, $m_{\rm min}=4.5 M_\odot$, $m_{\rm max}=100 M_\odot$, $\delta_m=4.8 M_\odot$, $\mu_g=35.5 M_\odot$, $\sigma_g=4.8 M_\odot$ and $\lambda_{\rm peak}=0.05$, which are consistent with the latest constraints from Ref.~\cite{KAGRA:2021duu}. 
We employ \texttt{gwfish} \cite{2023AnC....4200671D} to assess the detectability and the errors in the determination of the spin magnitude parameters for the simulated CBCs. Using \texttt{gwfish}, we calculate for each binary the matched filter SNR assuming $\Delta$-shaped Einstein Telescope with a sensitivity set to the case ``10 km HFLF cryogenic'' in \cite{2023JCAP...07..068B} and using the \texttt{IMRPhenomXPHM} waveform model. If the matched filter SNR exceeds a threshold of 12, we consider the binary to be detected and we approximate  the posterior on the spin magnitude measurement with a Fisher information matrix (FIM)~\footnote{Note that formally, the inverse of the FIM approximates the GW likelihood and not the posterior. In practice, one should use the GW likelihood (FIM) to draw an observed data point in the parameter space. Then, given the observed data point, one should calculate the GW posterior using FIMs that should be calculated on the entire parameter space. Consistently with \cite{2023arXiv231205302B}, we find that approximating the GW posterior with the FIM does not introduce any significant bias.}. We only consider a single configuration and sensitivity for the ET as the precision on the spin magnitudes does not strongly depend on it~\cite{2023JCAP...07..068B}.

\begin{figure}[t]
    \centering
    \includegraphics{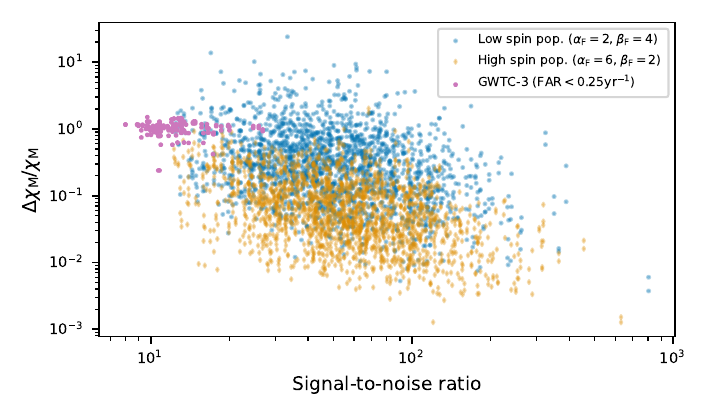}
    \caption{68.3\% credible interval fractional error on the spin magnitude (vertical axis) versus the source signal-to-noise ratio for simulated and real GW events. The blue circle and orange diamonds indicate the simulated binaries  for the low and high spin CBC populations observable with ET. The pink pentagons indicate the binaries observed during O3.}
    \label{fig:ET_distro}
\end{figure}

Figure \ref{fig:ET_distro} shows the typical precisions on the estimation of the spin magnitude parameters for BBH binaries observable with ET.  Given the enhanced sensitivity of ET, GW sources will be observed with significantly higher SNRs allowing for a more precise measurement of the spin magnitude than with current-generation GW detectors. From Fig.~\ref{fig:ET_distro}, we can also observe that the low spin population displays a lower precision in the spin magnitude (in terms of percentage error). The absolute precision on the spin magnitudes, $\Delta \chi_{\rm{M}}$, does not strongly depend on the value of the spin itself, but rather on the binary's signal-to-noise ratio. It follows that the low spin population has a worse precision in percentage error, $\Delta \chi_{\rm{M}}/\chi_{\rm{M}}$ as the values of $\chi_M$ are typically smaller.

\subsection{Forecasts}

We use three different population models for the forecast study: the \bbh, the \mixture (polar) and \mixture (axial) population models. For the three population models, we consider the same set of priors reported in Tab.~\ref{tab:priors}. We run all the forecast studies considering a set of 50, 100 and 200 GW detections with ET. For each GW detection, we provide a set of 4096 parameter estimation samples on the luminosity distance, detector frame masses and spin magnitudes generated with the Fisher information matrix. This choice corresponds to a maximum of $819200$ data points to use for the hierarchical Bayesian inference. ET will be able to collect $\mathcal{O}(100)$ of GW from BBHs in less than one day of observation\cite{Regimbau_2012, Regimbau_2014,Branchesi:2023mws}. The motivation for considering  $\mathcal{O}(100)$ of GW detections is that current codes for hierarchical Bayesian inference are too computationally demanding for $\mathcal{O}(>10^6)$ data points~\cite{Mastrogiovanni:2023zbw}.

\begin{figure}[t]
\centering
\begin{subfigure}{0.45\textwidth}
    \includegraphics[width=\textwidth]{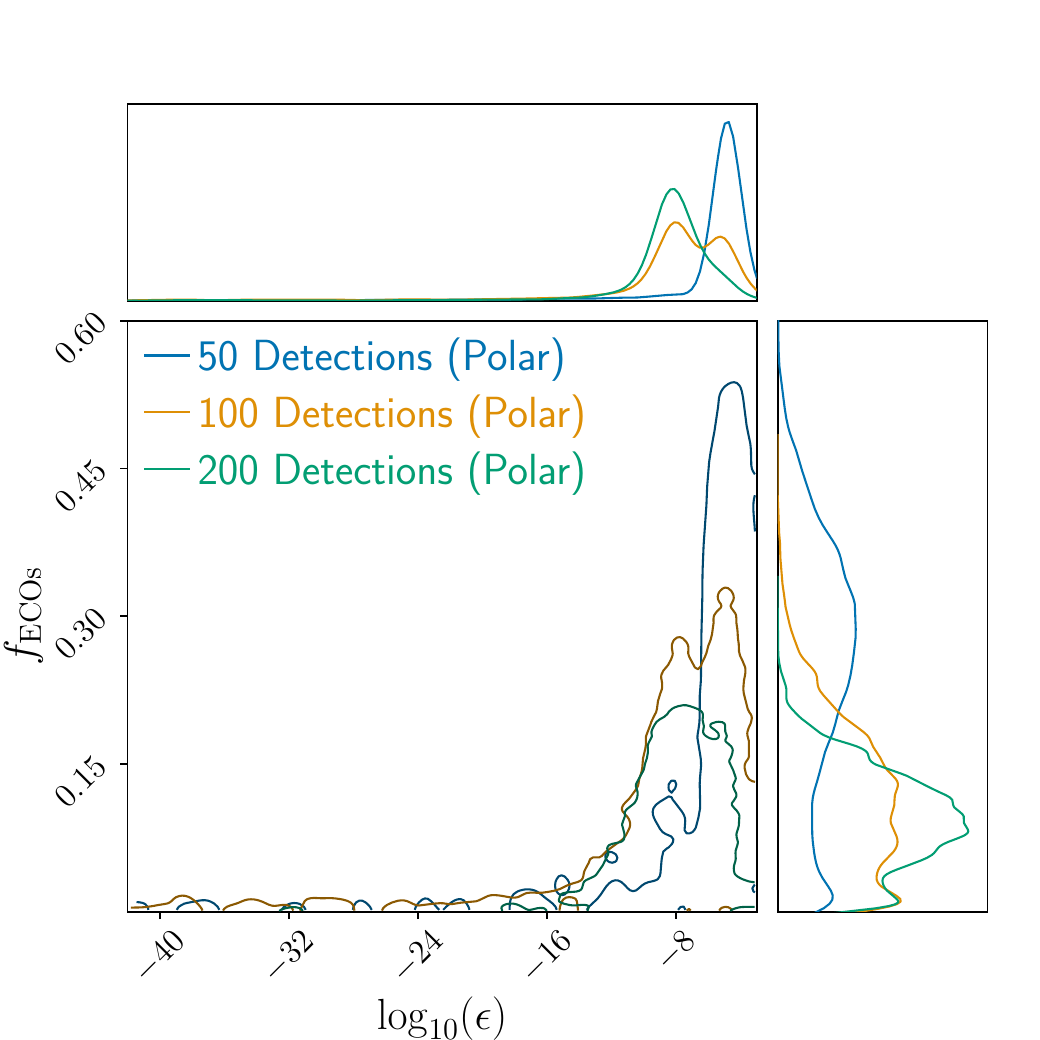}
    \caption{Low spin BBH population. Inference for ECOs with polar perturbations.}
    \label{fig:first}
\end{subfigure}
\hfill
\begin{subfigure}{0.45\textwidth}
    \includegraphics[width=\textwidth]{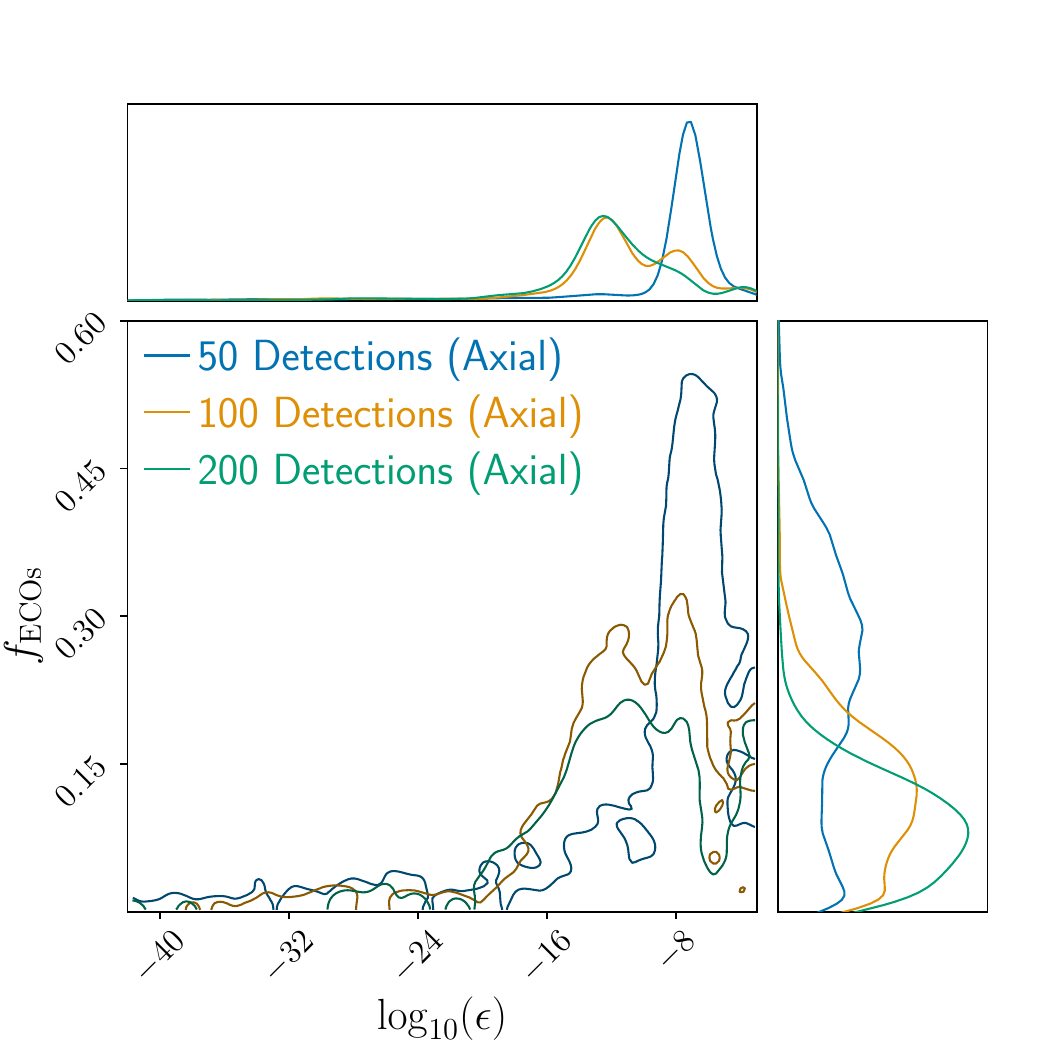}
    \caption{Low spin BBH population. Inference for ECOs with axial perturbations.}
    \label{fig:second}
\end{subfigure}
\hfill
\begin{subfigure}{0.45\textwidth}
    \includegraphics[width=\textwidth]{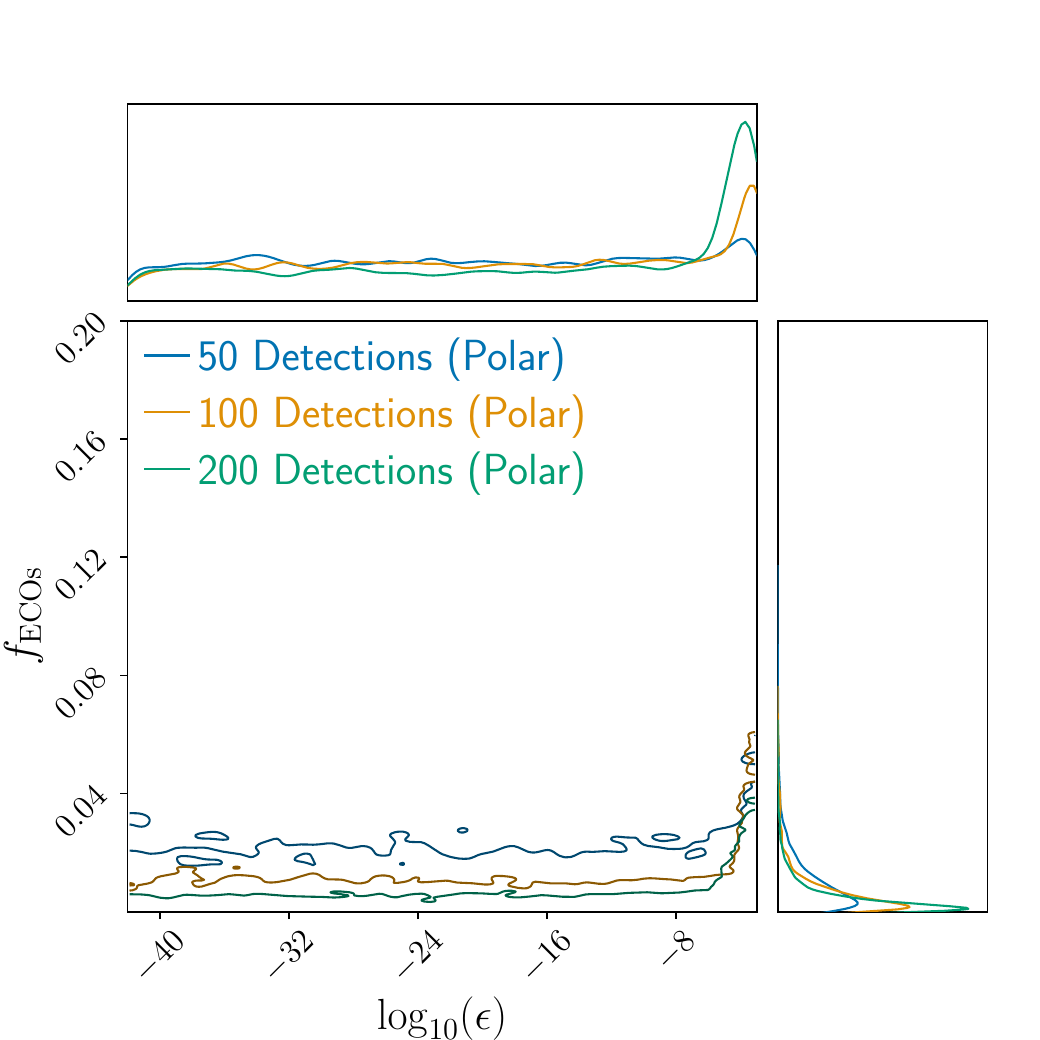}
    \caption{High spin BBH population. Inference for ECOs with polar perturbations.}
    \label{fig:third}
\end{subfigure}
\hfill 
\begin{subfigure}{0.45\textwidth}
    \includegraphics[width=\textwidth]{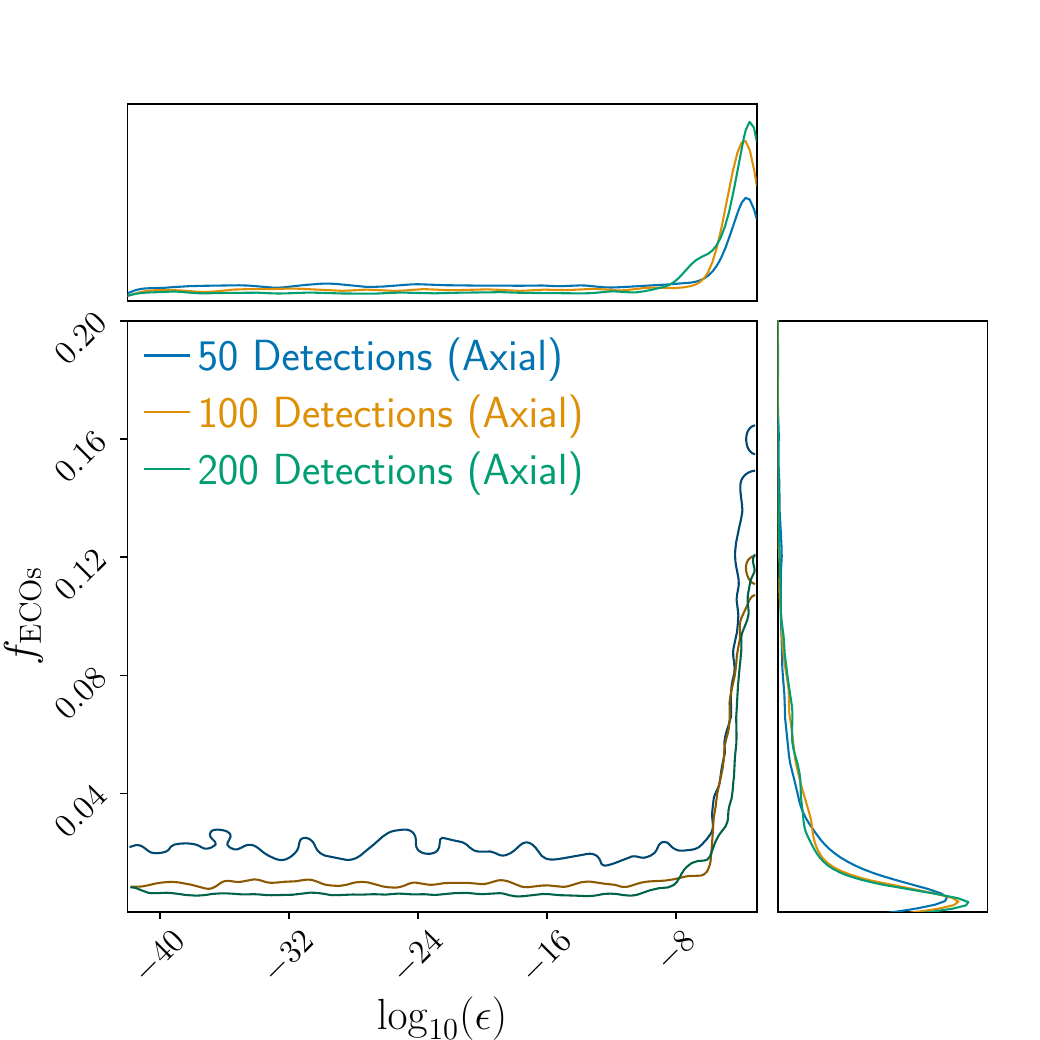}
    \caption{High spin BBH population. Inference for ECOs with axial perturbations.}
    \label{fig:fourth}
\end{subfigure}
\caption{Marginal posteriors on $\epsilon$ and \feco for the simulated low spin (top row) and high spin (bottom-row) BBH populations. The contours indicate the 90\% credible interval of the posteriors and the colors mark posteriors obtained with a different number of GW events.}
\label{fig:sims}
\end{figure}
\begin{figure}[t]
    \centering

    \includegraphics{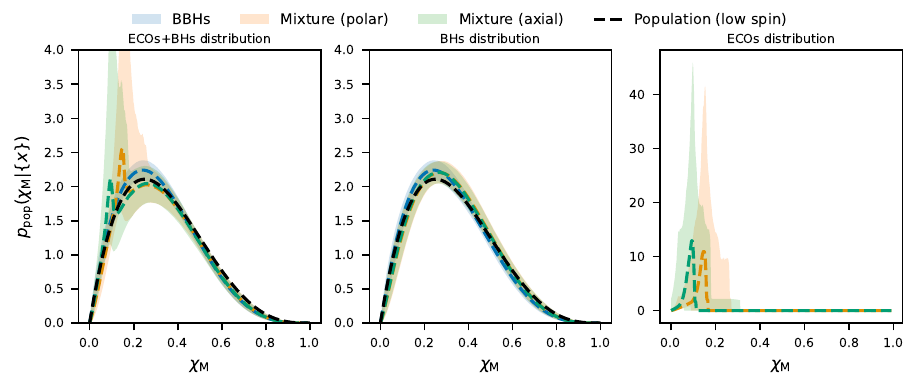}
    \includegraphics{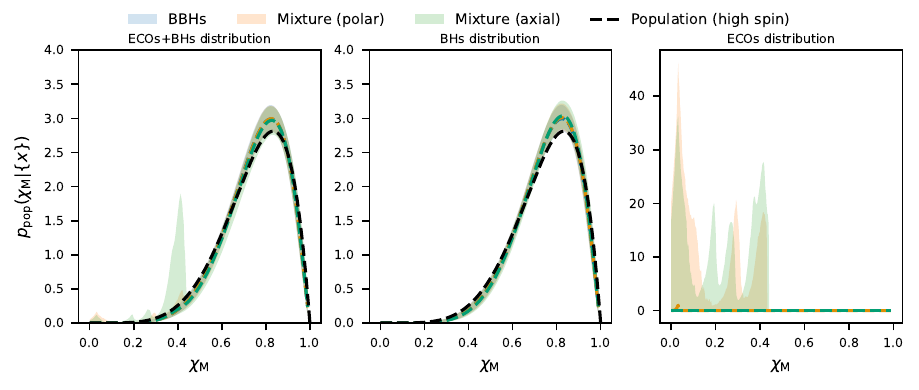}

    \caption{\textit{Left panels}: Spin magnitude distribution reconstructed for the overall population from the GW events detected during O3 for the BBHs-only and mixture cases. \textit{Middle panels}: Spin magnitude distribution reconstructed for the BBH population from the GW events detected during O3 for the BBHs-only and mixture cases. \textit{Right panels}: Spin magnitude distribution reconstructed for the ECOs population from the GW events detected during O3 for the mixture cases. The dashed lines indicate the simulated low spin (top panels) and high spin (bottom panels) populations and the shaded areas the 90\% C.I. reconstructed spin distributions.}
    \label{fig:ppcET}
\end{figure}

The overall upper limit on \feco is provided in Tab.~\ref{tab:ul} for the simulated populations. In Fig.~\ref{fig:sims} we report the 90\% credible intervals for the joint posterior on $\epsilon$ and \feco that could be obtained with ET, while in Fig.~\ref{fig:ppcET} we report the reconstructed spin distributions for the low and high spin populations. We find that a population of CBCs composed of slowly-spinning BBHs (as current data suggests) is worse in constraining the $\epsilon$-\feco parameter space if compared to a population of highly-spinning BBHs. This is because the maximum \chicrit allowed in the prior range of $\epsilon$ is 0.45 (0.68) for polar (axial) perturbations, which is high enough to include the bulk of BBHs produced at spins $\sim 0.2$ by the low spin population. As a consequence, higher values of $\epsilon$ and \feco can be used to fit a part of the BBH population at spin magnitudes around 0.2. This can be seen in Fig.~\ref{fig:sims} for the joint posterior, and in Fig.~\ref{fig:ppcET} for the spin reconstruction. From the point of view of \feco and $\epsilon$, the posterior supports higher values of \feco for $\epsilon \sim 10^{-8}$ as this region corresponds to a $\chicrit \sim 0.2$. From the point of view of the spin reconstruction, part of the population of the BBHs is reconstructed with an ECO population with a $\chicrit \sim 0.2$. The more the GW events, the stricter the constraints on the ECOs population.

The same considerations done for the low spin population can be made for the high-population with the different caveat that the majority of BBHs are produced with spins $\sim 0.8$. This value of the spin cannot be covered by the \chicrit predicted by $\epsilon$ that ranges to 0.45 (0.68) for polar (axial) perturbations. As a consequence, the posterior in Fig.~\ref{fig:sims} strongly constrains the values of $\epsilon$ and \feco, as  even a small fraction of ECOs produced with spins lower than $0.8$ would be distinguishable in the population's spin distribution. The fact that the population of ECOs allowed is very small can also be seen in Fig.~\ref{fig:ppcET} in which it is evident that the majority of the CBC population fits with a BBH population only.  

The low spin population produces lower constraints on \feco while the high spin population produces better constraints. With just one day of observation with ET, we will be able to constrain the fraction of ECOs present in the CBC population more accurately than what has been achieved with years of observation with current detectors. We do not report any upper/lower limit on $\epsilon$ as \feco=0 is a degenerate point that supports all the values of $\epsilon$ and therefore it is not possible to define an upper/lower limit. However, we note that smaller values of $\epsilon$ (ultra-compact objects) can be easily ruled out if we consider even a small fraction of ECOs in the population.

\section{Discussion and conclusions}
\label{sec:6}

This work aimed to find imprints of ECOs on the spin distribution of a population of coalescing compact binary objects with GW measurements, assuming that (possibly part of) the objects are perfectly reflecting exotic black-hole mimickers, which are prone to the ergoregion instability. So far, ECOs have mostly been probed by studying possible deviations in the ringdown of CBCs \cite{Cardoso:2016rao,Cardoso:2016oxy,Maggio:2020jml,Saketh:2024ojw} or searching for GW echoes \cite{Abedi_2017, Nakano:2017fvh, Westerweck_2018, Lo:2018sep, Maggio:2019zyv, Miani:2023mgl, Uchikata:2023zcu}.
\begin{table}
\centering
{
\small
\begin{tabular}[t]{||l|c|c|c|c||}
    \hline
    Events & High spin & Low spin & high spin & Low spin  \\
    & (Polar) & (Polar) & (Axial) & (Axial)  \\
    
    \hline
    50 & 0.03 & 0.44 & 0.09 & 0.44 \\
    100 & 0.03 & 0.25 & 0.08 & 0.25 \\ 
    200 & 0.02 & 0.17 & 0.08 & 0.18 \\
\hline 
\end{tabular}
    \caption{90\% credible level upper limits on the fraction of ECOs present in the CBC population from simulated ET detections. The first column reports the number of GW observations.}
    \label{tab:ul}
}
\end{table}
The first use of the ergoregion instability to set constraints on the population of horizonless compact objects is in Ref.~\cite{Barausse:2018vdb}. The authors used the stochastic gravitational-wave background (SGWB) produced by a potential population of ECOs to set upper limits on $\epsilon$. In particular, the non-detection of the SGWB during O1~\cite{LIGOScientific:2018mvr}, rules out Planckian models ($\epsilon \sim 10^{-40}$), or at least reduces to $50 \%$ the fraction of ECOs in a potentially mixed population of BHs and ECOs. In this work, by studying the resolved CBCs from GWTC-3, we set an upper limit from ECOs with Planckianian deviations ($\epsilon \sim 10^{-40}$), of  $f_{\rm{ECO}} < 0.25$ at $90 \%$ C.L. Additionally, we also set constraints for the fraction of ECOs to be present in the CBC population by varying the compactness up to $\epsilon \sim 10^{-3}$.

In this analysis, none of the models (BBHs only, ECOs only, mixture) is strongly favoured over the others. Highly spinning objects such as GW190517\_055101 and GW191109\_010717 are the only GW detections for which we can be sure that they are not ECOs. 
Our inability to comment on the nature of the slowly spinning GW events is due  to the fact that we are still not able to measure the spins at high precision, and we still do not have a large number of events, even less at relatively high SNRs. 

Let us note that possible systematics on the labelling as ECO or BH can be introduced by the choice of the population model. Several recent works have found tentative evidence for a correlation between the source masses and spin magnitude distributions \cite{Tiwari:2021yvr,Godfrey:2023oxb,Ray:2024hos,Pierra:2024fbl}. According to the literature, heavier compact objects than $40 M_\odot$ would prefer high spin magnitudes while lighter objects would prefer low spin magnitudes. By including these spin/mass models we expect to preferentially classify lighter compact objects with masses lower than $40 M_\odot$ as slowly spinning (and therefore ECOs), while higher mass objects would be classified as highly spinning and therefore BHs. As a consequence, the probabilities reported in Tab.~\ref{tab:tab_spin} are sensitive to possible correlations in the population model of masses and spins. However, our upper limit on $\epsilon$ and the fraction of ECOs in the population is mostly driven by the two GW events GW190517\_055101 and GW191109\_010717 that have masses $M \gtrsim 40 M_\odot$ and high spin magnitudes whose determination is not prior driven. Therefore, we do not expect the upper limit on $\epsilon$ to be strongly sensitive to cross-correlations between masses and spins in the population model.

The prospects should improve dramatically as long as third-generation detectors will become operational. Higher average SNRs will improve the precision on the spin measurements and consequently on the population studies. For example, if ultra-compact ECOs didn't exist ($\feco= 0$), we would constrain the fraction of ECOs with an upper limit of 2\% (17\%) at 95\% credible level for an high spin (low spin) population, with 200 binaries (only 1 day of observation) with the Einstein Telescope.

\subsection{Acknowledgments}
We are grateful to Konstantin Leyde for the internal review and feedbacks during the LVK internal review.
S.M. is supported by ERC Starting Grant No. 101163912–GravitySirens. 
E.M. is supported by the European Union’s Horizon Europe research and innovation programme under the Marie Skłodowska-Curie grant agreement No. 101107586.
E.M. acknowledges funding from the Deutsche Forschungsgemeinschaft (DFG) - project number: 386119226. 
This material is based upon work supported by NSF's LIGO Laboratory which is a major facility fully funded by the National Science Foundation.
The authors are grateful for computational resources provided by the LIGO Laboratory and supported by National Science Foundation Grants PHY-0757058 and PHY-0823459.
This research has made use of data or software obtained from the Gravitational Wave Open Science Center (gwosc.org), a service of LIGO Laboratory, the LIGO Scientific Collaboration, the Virgo Collaboration, and KAGRA. LIGO Laboratory and Advanced LIGO are funded by the United States National Science Foundation (NSF) as well as the Science and Technology Facilities Council (STFC) of the United Kingdom, the Max-Planck-Society (MPS), and the State of Niedersachsen/Germany for support of the construction of Advanced LIGO and construction and operation of the GEO600 detector. Additional support for Advanced LIGO was provided by the Australian Research Council. Virgo is funded, through the European Gravitational Observatory (EGO), by the French Centre National de Recherche Scientifique (CNRS), the Italian Istituto Nazionale di Fisica Nucleare (INFN) and the Dutch Nikhef, with contributions by institutions from Belgium, Germany, Greece, Hungary, Ireland, Japan, Monaco, Poland, Portugal, Spain. KAGRA is supported by Ministry of Education, Culture, Sports, Science and Technology (MEXT), Japan Society for the Promotion of Science (JSPS) in Japan; National Research Foundation (NRF) and Ministry of Science and ICT (MSIT) in Korea; Academia Sinica (AS) and National Science and Technology Council (NSTC) in Taiwan. 

\appendix

\section{The rate model and priors used for the agnostic run} \label{app:A}

In this Appendix, we specify the various distributions and analytical functions that are defined to build the CBC merger rate used for this paper.
As mentioned in Sec.~\ref{sec:stat_model}, the CBC rate model is parametrized as
\begin{eqnarray}
    \frac{\de \Ncbc}{\de \mos \de \mts \de \com \de \ctm \de z \, \de \Omega \, \de t_s} (\Lambda) &=& R_0 \psi(z;\Lambda_r) \frac{\de V_c}{\de z \, \de \Omega}  p_{\rm pop}(\mos,\mts|\Lambda_m) \times \nn \\ &&  p_{\rm pop}(\com|\Lambda_s)p_{\rm pop}(\ctm|\Lambda_s),
    \label{eq:allrate}
\end{eqnarray}
where $R_0$ is the rate of events per comoving volume at the present time, and $\psi(z;\Lambda_r)$ parametrizes the dependence of the CBC merger rate on redshift and $V_c$ is the comoving volume. The $p_{\rm pop}(\cdot)$ functions are the population (astrophysical) distributions of the objects' source masses and spins. The spin magnitude distribution is constructed for a population of ECOs and classical BBHs in Sec.~\ref{sec:pop_model}, and the priors used for the models are highlighted in Tab.~\ref{tab:priors}.

We model the CBC rate evolution following a Star-Formation-like parametrization \cite{Madau:2014bja}
\begin{equation}
    \psi(z;\gamma)=[1+(1+z_p)^{-\gamma-k}] \frac{(1+z)^\gamma}{1+\left(\frac{1+z}{1+z_p}\right)^{\gamma+k}}\,.
    \label{eq:ratemod2}
\end{equation}
where $\gamma, k$ and $z_p$ are the population parameters governing the CBC merger rate at $z \lesssim z_p$ and $z \gtrsim z_p$. The source masses are modelled as follows,
\begin{eqnarray}
p_{\rm pop}(m_{1,s},m_{2,s}|\Lambda)=\pi(m_{1,s}|\Lambda)\pi(m_{2,s}|m_{1,s},\Lambda)] \times \nonumber S(m_1|m_{\rm min},\delta_m)S(m_2|m_{\rm min},\delta_m),
\label{massprior}
\end{eqnarray}
with 
\begin{eqnarray}
\label{eq:smoothing}
&& S(m_{1,s} | m_{\rm min}, \delta_m) = \left\{
\begin{array}{lr}
    0, & \left(m< m_{\rm min}\right) \\
    \left[f(m - m_{\rm min}, \delta_m) + 1\right]^{-1}, & \quad \left(m_{\rm min} \leq m < m_{\rm min}+\delta_m\right) \\
    1, & \quad \left(m\geq m_{\rm min} + \delta_m\right) 
\end{array}
\nonumber \right. \\
\end{eqnarray}
defined as a logistic-like window function that introduces a smooth transition between 0 and 1, governed by the parameter $\delta_m$, starting from the minimum of the mass spectrum at $m_{\rm min}$. The transition function is given by
\begin{equation}
    f(m', \delta_m) = \exp \left(\frac{\delta_m}{m'} + \frac{\delta_m }{m' - \delta_m}\right).
    \label{eq:deltam}
\end{equation}
The mass distributions are modeled as 
\begin{eqnarray}
    \pi(m_{1,s}|m_{\rm min},m_{\rm max},\alpha)&=&(1-\lambda)\mathcal{P}(m_{1,s}|m_{\rm min},m_{\rm max},-\alpha)+ \nonumber \lambda \mathcal{G}(m_{1,s}|\mu_g,\sigma)\,, \\ &&\quad (0 \leq \lambda\leq 1) \,, \nn \\
    \pi(m_{2,s}|m_{\rm min},m_{1,s},\beta)&=&\mathcal{P}(m_{2,s}|m_{\rm min},m_{1,s},\beta)\, ,
\end{eqnarray}
where $\mathcal{P}(\cdot)$ indicates a Power Law distribution starting at $m_{\rm min}$ and stopping at $m_{\rm max}$ with slope parameter $\alpha$, while $\mathcal{G}(\cdot)$ is a gaussian distribution with mean $\mu_{g}$ and standard deviation $\sigma_g$. The parameter $\lambda$ is the fraction of the CBC population with masses in the gaussian part of the spectrum. The secondary mass is modeled enforcing the condition $m_{1,s}>m_{2,s}$ as a truncated power law with slope parameter $\beta$.
The priors used for the full population run are reported in Tab.~\ref{tab:priors_full}.

\begin{table}
\centering
\begin{tabular}[t]{|c c|}
 \hline
 Population parameter & Prior \\
 \hline\hline
 Primary mass power law slope $\alpha$ & $\mathcal{U}[1.5,12]$ \\
 Secondary mass power law slope $\beta$ & $\mathcal{U}[-4,12]$ \\
 Minimum mass spectrum mass $m_{\rm min}$ & $\mathcal{U}[2,10]$ \\
 Maximum mass spectrum mass $m_{\rm max}$ & $\mathcal{U}[50,100]$ \\
 Transition scale at low masses $\delta_m$ & $\mathcal{U}[0,10]$ \\
 Primary mass gaussian peak $\mu_g$ & $\mathcal{U}[20,50]$ \\
 Primary mass gaussian standard deviation $\sigma_g$ & $\mathcal{U}[0.4,10]$ \\
 Fraction of sources in gaussian peak $\lambda$ & $\mathcal{U}[0,1]$ \\ 
 Rate evolution parameter, low redshift $\gamma$ & $\mathcal{U}[0,12]$ \\
Rate evolution parameter, high redshift $k$ & $\mathcal{U}[0,6]$ \\ 
Peak of CBC merger rate $z_p$ & $\mathcal{U}[0,4]$ \\
 \hline
 \end{tabular}
 \caption{Priors used on the analysis for the full population parameters run.}
\label{tab:priors_full}
\end{table}

\section{Computation of the hyperposterior} \label{app:B}

The starting point for the reconstruction of the hyperposterior $p(\Lambda|\{x\})$ is, given a point in the hyperparameter space $\Lambda_0$, the computation of the hyperlikelihood $\mathcal{L}(\{x\}|\Lambda_0)$, that we rewrite for clearness,
\begin{equation}
    \mathcal{L}(\{x\}|\Lambda) \propto  e^{-N_{\rm exp}(\Lambda)} \prod_{i=1}^{\Ngw} \Tobs \int \de \theta  \de z \; \mathcal{L}_{\rm GW}(x_i|\theta,z)  \times \frac{1}{1+z} \frac{\de \Ncbc}{\de \theta \de z \de t_s}(\Lambda),
    \label{eq:fund_2}
\end{equation}
As stated in the main text, the term \Nexp takes into account the selection effect, due to the finite sensitivity of the detectors, and can be written in general as
\begin{equation}
    \Nexp (\Lambda)= \Tobs \int \de \theta \de z \; \Pdet(\theta, z) \, \frac{1}{1+z} \frac{\de \Ncbc}{\de z  \de \theta \de t_s}(\Lambda) ,
    \label{eq:Nexp}
\end{equation}
 where the detection probability $\Pdet(z,\theta)$ represents the probability that an event characterized by true parameters $\theta$ at redshift $z$ is detected, i.e. it overpasses some chosen detection threshold adopted by the search algorithm (e.g. the SNR or the false alarm rate, FAR). It is defined by integrating the likelihood over all the possible data realizations that are detectable, namely
 \begin{equation}
     \Pdet(\theta, z) = \int_{x \in {\rm detectable}} \de x \; \mathcal{L}_{\rm GW}(x|\theta,z).
 \end{equation}
 
To evaluate Eq.~(\ref{eq:fund_2}), we implement the rate model in Eq.~(\ref{eq:allrate}) inside the python package \textsc{icarogw}~\cite{Mastrogiovanni:2023zbw}. \textsc{icarogw} is a software developed for hierarchical Bayesian inference able to calculate the likelihood in Eq.~(\ref{eq:fund_2}) starting from a set of ``posterior samples'' from detected GW events and an injection set generated to account for selection effects. In particular, the integrals of Eq.~(\ref{eq:fund_2}) can be computed by noting that, using Bayes theorem, the GW likelihoods can be recast in the form 
\begin{equation}
\mathcal{L}_{\rm GW}(x_i|\theta,z) \propto \frac{p(\theta,z|x_i)}{\pi_{\rm PE}(\theta, z)} ,
\end{equation}
where $p(\theta,z|x_i)$ is the posterior distribution for the $i$-th detected event of the population, while $\pi_{\rm PE}(\theta, z)$ is the prior distribution related to the single-event parameter estimation. The Bayesian evidence $p(x_i)$ is not function of the parameters $(\theta, z)$, and can be absorbed in the proportionality coefficients of Eq.~(\ref{eq:fund_2}). Having a set of posterior samples of each of the events $x_i$, one can approximate the aforementioned integrals with a Monte Carlo technique,
\begin{eqnarray}
&&\int \de \theta  \de z \, \mathcal{L}(x_i|\theta,z,\Lambda) \frac{1}{1+z} \frac{\de N}{ \de\theta \de z \de t_s}(\Lambda)  \approx \nn \\ &&\frac{1}{N_{{\rm s},i}} \sum_{j=1}^{N_{{\rm s},i}} \frac{1}{\pi_{\rm PE}(\theta_{i,j}, z_{i,j}|\Lambda)}\frac{1}{1+z_{i,j}}\frac{dN}{\de t_s \de \theta \de z}(\Lambda)\bigg|_{i,j} 
     \label{eq:intpe}
\end{eqnarray}
where the index $i$ refers to the event, while $j$ runs on the $N_{s,i}$ posterior samples.

The evaluation of the selection effects, encoded in Eq.~(\ref{eq:Nexp}), is performed similarly. Firstly, a set of ``injection'' events, characterized by parameters ($\theta_j, z_j$), is extracted from a certain distribution $\pi_{\rm inj }(\theta, z)$. Then, rearranging the formula as
\begin{equation}
    \Nexp (\Lambda)= \Tobs \int \de \theta \de z \; \pi_{\rm inj}(\theta, z) \Pdet(\theta, z) \left[ \frac{1}{\pi_{\rm inj}(\theta, z)} \, \frac{1}{1+z} \frac{\de \Ncbc}{\de t_s \de z  \de \theta}(\Lambda) \right] \,,
\end{equation}
one can employ the Monte Carlo approach, with the selection term $\Pdet(z,\theta)$ that has the effect of counting only the $N_{det}$ events above the selected threshold. Therefore we have
\begin{equation}
    N_{\rm exp}(\Lambda)  \approx \frac{T_{\rm obs}}{N_{\rm gen}} \sum_{j=1}^{N_{\rm det}} \frac{1}{\pi_{\rm inj}(\theta_j, z_j)}\frac{1}{1+z_j}\frac{dN}{dt_s dz d\theta} (\Lambda) \bigg|_j .
    \label{eq:nexpnum}
\end{equation}

Putting pieces together, the code approximates the hierarchical likelihood as 
\begin{equation}
    \ln[\mathcal{L}(\{x\}|\Lambda)] \approx -\frac{T_{\rm obs}}{N_{\rm gen}} \sum_{j=1}^{N_{\rm det}} s_j + \sum_{i}^{N_{\rm obs}} \ln\left[ \frac{T_{\rm obs}}{N_{{\rm s},i}} \sum_{j=1}^{N_{{\rm s},i}} w_{i,j} \right]\,,
        \label{eq:hl_numer}
\end{equation}
where $s_j$ and $w_{i,j}$ are the approximated integrals, where the $i$ index refers to the $i$-th event, while $j$ to the Monte Carlo sampling:
\begin{eqnarray}
    s_j & = & \frac{1}{\pi_{\rm inj}(\theta_j, z_j)}\frac{1}{1+z_j}\frac{dN}{dt_s dz d\theta}(\Lambda)\bigg|_j \,, \\
    w_{i,j} & = & \frac{1}{\pi_{\rm PE}(\theta_{i,j}, z_{i,j}|\Lambda)}\frac{1}{1+z_{i,j}}\frac{dN}{\de t_s \de \theta \de z}(\Lambda)\bigg|_{i,j} \,.
\end{eqnarray}
Having a way to compute the hyperlikelihood and given some prior distribution $p(\Lambda)$, the package can generate posterior samples exploiting Monte Carlo Markov Chain algorithms~\cite{2021MNRAS.507.2037A, Foreman_Mackey_2013} or nested samplers~\cite{Speagle_2020}. For this purpose, \icarogw is interfaced with \texttt{Bilby}~\cite{Ashton_2019}, a Python package for gravitational-wave inference and data analysis. \\

\bibliographystyle{ieeetr}
\bibliography{refs}

\end{document}